\documentclass[lettersize,journal]{IEEEtran}
\usepackage{amsmath,amsfonts}
\usepackage{textcomp}
\usepackage{stfloats}
\usepackage{url}
\usepackage{verbatim}
\usepackage{graphicx}

\usepackage{cite}
\usepackage{textcomp}
\usepackage{graphicx}
\usepackage[shortlabels]{enumitem}
\usepackage{balance}
\usepackage{comment}

\usepackage{url}
\usepackage{color}

\usepackage{amsmath,amsfonts,amssymb}
\usepackage{bm}
\usepackage{algorithm}
\usepackage{algpseudocode}

\usepackage{wasysym}

\usepackage{multirow}
\usepackage[numbers]{natbib}
\usepackage{blindtext}
\usepackage{wrapfig}
\usepackage{booktabs}
\usepackage{subfigure}

\usepackage{tcolorbox}
\tcbuselibrary{breakable}
\usepackage{xcolor}
\usepackage{soul}

\usepackage{colortbl}

\definecolor{lightpink}{rgb}{1.0, 0.87, 0.87}
\definecolor{lightpurple}{rgb}{0.94, 0.85, 0.94}
\definecolor{lightgreen}{rgb}{0.88, 1.0, 0.88}
\definecolor{lightyellow}{rgb}{1.0, 1.0, 0.88}
\definecolor{lightblue}{rgb}{0.88, 0.94, 1.0}
\definecolor{lightgrey}{rgb}{0.8, 0.8, 0.8}

\usepackage{listings}

\definecolor{codegreen}{rgb}{0,0.6,0}
\definecolor{codegray}{rgb}{0.5,0.5,0.5}
\definecolor{codepurple}{rgb}{0.58,0,0.82}
\definecolor{backcolour}{rgb}{0.95,0.95,0.92}

\lstdefinestyle{mystyle}{
    backgroundcolor=\color{backcolour},   
    commentstyle=\color{codegreen},
    keywordstyle=\color{magenta},
    numberstyle=\tiny\color{codegray},
    stringstyle=\color{codepurple},
    basicstyle=\footnotesize\ttfamily,
    breakatwhitespace=false,         
    breaklines=true,                 
    captionpos=b,                    
    keepspaces=true,                 
    numbers=left,                    
    numbersep=5pt,                  
    showspaces=false,                
    showstringspaces=false,
    showtabs=false,   
    tabsize=3,
    frame=tb,
    aboveskip=2mm,
    belowskip=2mm
}

\usepackage{hyperref}
\usepackage[hyphenbreaks]{breakurl}

\newcounter{DaveCommentCounter}
\newcommand{\ddu}[1]{
    \stepcounter{DaveCommentCounter}
    \textcolor{blue}{\textit{/**Dave's comment [\arabic{DaveCommentCounter}]: I don't understand the intended meaning in the next sentence. Please revise/delete/explain. **/}}
}

\newcommand{\dns}[1]{
    \stepcounter{DaveCommentCounter}
    \textcolor{blue}{\textit{/**Dave's comment [\arabic{DaveCommentCounter}]: I'm not sure that I have captured the intended meaning in the next sentence. Please check/confirm.. **/}}
}

\newcounter{ZixiangCommentCounter}
\newcounter{RubingCommentCounter}
\begin{document}

\title{TransformCode: A Contrastive Learning Framework for Code Embedding via Subtree Transformation}

\author{Zixiang Xian, Rubing Huang,~\IEEEmembership{Senior Member,~IEEE,} Dave Towey,~\IEEEmembership{Senior Member,~IEEE,} Chunrong Fang,~\IEEEmembership{Member,~IEEE,} Zhenyu Chen,~\IEEEmembership{Senior Member,~IEEE}

        \thanks{Zixiang Xian and Rubing Huang are with the School of Computer Science and Engineering, Macau University of Science and Technology, Taipa, Macau 999078, China. E-mail: 3220001352@student.must.edu.mo, rbhuang@must.edu.mo.}

        \thanks{Dave Towey is with the School of Computer Science, University of Nottingham Ningbo China, Ningbo, Zhejiang 315100, China. E-mail: dave.towey@nottingham.edu.cn.}

        \thanks{Chunrong Fang and Zhenyu Chen are with the State Key Laboratory for Novel Software Technology, Nanjing University, Nanjing 210093, China, E-mail: \{fangchunrong, zychen\}@nju.edu.cn.}
        \IEEEcompsocitemizethanks{\IEEEcompsocthanksitem Rubing Huang and Chunrong Fang are the corresponding authors.}
}

\maketitle

\begin{abstract}
Artificial intelligence (AI) has revolutionized software engineering (SE) by enhancing software development efficiency.
The advent of pre-trained models (PTMs) leveraging transfer learning has significantly advanced AI for SE.
However, existing PTMs that operate on individual code tokens suffer from several limitations: They are costly to train and fine-tune; and they rely heavily on labeled data for fine-tuning on task-specific datasets
In this paper, we present \textbf{TransformCode}, a novel framework that learns code embeddings in a contrastive learning manner. 
Our framework is encoder-agnostic and language-agnostic, which means that it can leverage any encoder model and handle any programming language. 
We also propose a novel data-augmentation technique called \textbf{abstract syntax tree (AST) transformation}, which applies syntactic and semantic transformations to the original code snippets, to generate more diverse and robust samples for contrastive learning. Our framework has several advantages over existing methods: 
(1) It is flexible and adaptable, because it can easily be extended to other downstream tasks that require code representation (such as code-clone detection and classification); 
(2) it is efficient and scalable, because it does not require a large model or a large amount of training data, and it can support any programming language; 
(3) it is not limited to unsupervised learning, but can also be applied to some supervised learning tasks by incorporating task-specific labels or objectives; and 
(4) it can also adjust the number of encoder parameters based on computing resources.
We evaluate our framework on several code-related tasks, and demonstrate its effectiveness and superiority over the state-of-the-art methods such as SourcererCC, Code2vec, and InferCode.
\end{abstract}


\section{Introduction}

Over the past decade, artificial intelligence (AI) has gained immense popularity within the field of software engineering (SE). The integration of AI into SE practices primarily aims to enhance the efficiency and effectiveness of software development processes, thereby enhancing productivity and fostering innovation. The application of AI to SE has been notably enhanced by the advent of pre-trained models (PTMs) \cite{codebert2020,graphcodebert2021}. Trained on extensive code datasets, PTMs exhibit a profound comprehension of programming languages, enabling them to be fine-tuned for various SE tasks \cite{codebert2020,graphcodebert2021,ahmad-etal-2021-unified}. The true strength of these PTMs lies in their capacity for transfer learning. This is a process in which a model, having acquired knowledge from one task, applies this learned knowledge to a different but related task. The use of transfer learning effectively capitalizes on the vast, pre-existing knowledge embedded within the PTMs, providing a rich, pre-trained context for code interpretation. 

The emergence of PTMs has significantly advanced the domain of code embedding. CodeBERT \cite{codebert2020} has shown good results on a variety of code-related tasks. This model is distinguished by its use of the Transformer architecture, a novel neural network design that employs self-attention mechanisms \cite{vaswani2017attention}. These mechanisms excel in capturing complex global dependencies between inputs and outputs within specific contexts. Therefore, they can enhance the model's ability to learn code embedding.
Code embedding \cite{bui2023codetf} involves converting source code into a format that machine learning models can handle, such as vectors.
Hence, PTMs, particularly Transformers, have been used to address various SE tasks, such as code intelligence \cite{bui2023codetf}, sentiment analysis \cite{batra2021bert}, code summarization \cite{gaoCodeStructureGuided2022,ahmad2020transformer,tangASTtransCodeSummarization2022,zhangRetrievalbasedNeuralSource2020,liuRetrievalAugmentedGenerationCode2022}, and vulnerability detection \cite{wuVulCNNImageinspiredScalable2022, chakrabortyDeepLearningBased2020}.
Studies have shown the impressive capability of Transformers for mapping and generating natural language and source code.
They have also introduced some novel and improved methods, such as incorporating code-structure information, integrating multiple models, compressing models, and combining other features.
PTMs are capable of performing various SE tasks, but they can also require huge GPU resources and high-quality datasets for fine-tuning, due to their massive numbers of parameters.
They are usually pre-trained on large-scale datasets and then fine-tuned on task-specific datasets.
However, in many cases, it may be preferable to use unsupervised or self-supervised learning to train a smaller model directly on our own dataset, without requiring on too many other resources.

This paper proposes a novel framework for learning code embeddings from arbitrary datasets, which can be leveraged for various SE tasks.
\textbf{To the best of our knowledge, this is the first work that applies contrastive learning to code embeddings by comparing the original abstract syntax tree (AST) of the code with its transformed version.}

Our proposed framework combines contrastive learning with a momentum encoder \cite{he2020momentum,chen2020improved} to learn powerful and robust representations from unlabeled data. 
Contrastive learning aims to maximize the similarity between an encoded query and a positive key, while minimizing the similarity with negative keys 
(where a query is a training sample; a positive key is a transformed version of that sample; and a negative key is a sample from a different input). 
We use a queue and a momentum-updated encoder to construct a dynamic and consistent dictionary of keys, which allows us to perform contrastive learning with a large and diverse set of negative samples. 
Our framework improves contrastive learning by incorporating additional features of transformed code that enhance the quality and efficiency of the learned representations. 
Furthermore, we combine contrastive loss with supervised labels for the supervised learning tasks.

Our main contributions in this paper are as follows:
\begin{itemize}
        \item[(1)]
            We first propose a novel framework that uses contrastive learning to learn code embeddings from unlabeled code. 
            Although our proposed framework is encoder-independent, we use the Transformer encoder, and modify its positional encoding to better capture the structure of the code. 
            The encoder's architecture can be tuned based on the available computing resources.              
        \item[(2)]
            We propose a pipeline for contrastive learning of code that consists of three steps: 
            code normalization; 
            code transformation; and training. We design various \textbf{code transformation methods} to generate anchor samples that are semantically equivalent to the original code, but have different syntactic forms or structures. 
            These anchor samples are then used to train a contrastive learning model that can capture the semantic similarity of the code.

        \item[(3)]
            We present a state-of-the-art framework that can handle unlabeled real-world datasets with high efficiency and scalability. 
            Our framework enhances the performance by training customized tokenizers for each dataset, unlike existing methods that rely on generic tokenizers. Our customized tokenizers use a smaller and more efficient dictionary that does not require any special tokens.
            This enables us to handle the diversity and specificity of code snippets in various programming languages and tasks.
            Moreover, our framework can adapt to use different amounts of computational resources by adjusting the number of parameters used in the encoder.

        \item[(4)]
            Our proposed framework can achieve good performance on both unsupervised and supervised SE tasks.
            We evaluate our framework on several code-related tasks and demonstrate its effectiveness and superiority over existing methods (such as SourcererCC, Code2vec, and InferCode).
\end{itemize}

Compared with other frameworks, our proposed framework has several advantages over existing methods for learning code representation:
\begin{itemize}
    \item
        We propose a novel framework for learning code representations that is highly flexible and adaptable. 
        Unlike previous methods that rely on labeled data, our framework uses contrastive learning to capture the semantic similarity of code snippets. 
        This enables the framework to handle a wide range of downstream tasks that require code representation (such as detecting and classifying code clones, as well as clustering code snippets in an unsupervised manner).
        For example, we can apply model selection and a Gaussian-based mixture model \cite{xian_2021_AGGM,xian_2021_MML,Xian2022} to cluster code snippets with the code embedding learnt by our framework, without any labels.

    \item
        Our framework is very efficient and scalable across different programming languages. 
        Unlike existing methods that require a large model size or a large amount of training data, our framework can learn code representations from a small amount of data by using contrastive learning and data-augmentation techniques. 
        Our framework uses a self-attention mechanism that can dynamically adjust the importance of different nodes and edges in the code graph, and a contrastive learning objective that can learn from unlabeled code snippets.

    \item
        Our framework is a versatile tool for code-representation learning that can handle both unsupervised and supervised learning scenarios. 
        By incorporating task-specific labels or objectives into the contrastive-learning objective, our framework can leverage supervised signals to learn more discriminative and relevant code representations. 
        This makes it possible to address various supervised learning tasks (such as code summarization, code completion, and code-defect detection).

    \item
        Our framework can adjust the number of the encoder parameters according to the available computing resources, making it suitable for various deployment scenarios. We evaluated our framework on a range of SE tasks, using different programming languages and datasets, to demonstrate its effectiveness and robustness. 
        The results confirm that our framework can achieve comparable or better performance than the state-of-the-art methods, and can adapt well to different domains and languages.
\end{itemize}

The rest of this paper is organized as follows:
Section \ref{sec_related} reviews the related work and background about source-code embedding.
Section \ref{sec_code_to_ast} describes the process of transforming code snippets into ASTs that can be used as input for our framework.
Section \ref{sec_framework} presents our unsupervised learning framework for code embedding, and explains how it employs contrastive learning to learn meaningful representations.
Section \ref{sec_experiment} reports the results of our experiments on both supervised and unsupervised learning for several SE tasks, and compares our framework with other unsupervised code-embedding methods, showing its superior performance and effectiveness.
Finally, Section \ref{sec_conclusion} summarizes our contributions and outlines possible directions for future work.

\section{Related Work
  \label{sec_related}}

Code-embedding learning is an important task for various SE applications. The goal is to encode source code into vectors that capture its semantics and structure. 
A typical application is code-clone detection \cite{fangFunctionalCodeClone2020, mehrotraModelingFunctionalSimilarity2020, liuCanNeuralClone2021}, which aims to find code fragments that are similar or identical in functionality or syntax. 
These code clones may imply code-quality issues, such as redundancy, plagiarism, or inconsistency. 
By transforming code into vector representations, we can measure the similarity between code snippets, and determine whether or not they are clones.

\subsection{Methodology for Code Embedding}

Code-embedding methods can represent software code in a vector space using different types of input data for training. 
These methods use different types of input data --- such as plain text, syntax trees, or graphs --- to train their models and learn the semantic properties of the code. 
In this paper, we survey these methods and classify them into three main categories based on the form of code data that they use: token-based, tree-based, and graph-based approaches. 
Token-based methods treat code as a sequence of lexical tokens or n-grams as the basic units.
Tree-based methods parse code into abstract syntax trees (ASTs) or other tree structures that capture the syntactic and semantic rules of code.
Graph-based methods construct graphs from code, such as control flow graphs (CFGs), data flow graphs (DFGs), or other graph structures that represent the dynamic behavior and dependencies of code.

One way to extract code features using a token-based approach is to apply Term Frequency-Inverse Document Frequency (TF-IDF), which measures the importance of each word in a document based on its frequency. 
These features can then be used as inputs for discriminative models such as Support Vector Machines (SVMs) \cite{svm_1998} or Extreme Gradient Boosting (XGBoost) \cite{Chen_2016_XGBoost}, which learn to classify or predict labels from the data. However, this method has a major drawback: 
It ignores the syntax and semantics of programming languages, which are crucial for understanding code. 
To address this issue, researchers have investigated various deep learning methods that can learn more rich and meaningful representations of code. 
One of the popular token-based deep learning methods involves tokenizing code and training a language model like BERT using Masked Language Modeling (MLM) and Next Sentence Prediction (NSP) \cite{devlin2018bert}, which is the core idea of CodeBERT \cite{codebert2020}.
CodeAttention \cite{zheng_codeattention_2019} is another token-based method that translates source code to natural language comments by leveraging the code structures. 
These methods extract keywords or topics from various sources of information, such as the API knowledge embedded in the code, the document description provided by the developers, or the identifier naming conventions, etc. 
Ahmad et al. \cite{ahmad2020transformer} proposed a simple yet effective token-based Transformer model that employs relative position encoding and copy attention to generate natural language summaries of source code.

Code-embedding approaches based on trees and graphs aim to improve on token-based approaches' capture of the semantic and structural aspects of code. 
Tree-based code embedding relies on the fact that code follows strict syntactic rules that can be parsed as ASTs (which are tree-like representations of the syntactic structure of code). 
However, tree-based code embedding also needs a suitable neural network architecture to handle the tree-shaped data. 
One possible architecture based on tree-based approaches is the tree-based neural network \cite{Ye2023tree_based}, which is a neural network that resembles the structure of a decision tree or a recursive tree. 
Mou et al. \cite{mouConvolutionalNeuralNetworks2015} proposed a tree-based convolutional neural network (TBCNN) that uses ASTs to encode source code snippets and perform various program analysis tasks, such as functionality classification and pattern detection. 
Zhang et al. \cite{zhangNovelNeuralSource2019} developed ASTNN, a novel neural network model that can represent source code fragments based on ASTs: 
The model first splits each code fragment into a sequence of statement trees, which are then encoded by a tree-based neural network. 
Then the model uses a bidirectional recurrent neural network (RNN) to capture the naturalness of the statements and generate the final vector representation of the code fragment.
Alon et al. \cite{alon2019code2vec} introduced code2vec, which can represent source code fragments as fixed-length vectors (code embeddings) by decomposing them into a set of paths in their ASTs and learning how to aggregate them. 
Alon et al. \cite{alon2018code2seq} later extended code2vec to code2seq by using long short-term memory networks (LSTMs) to enable variable-length sequences and capture more syntactic information with multiple AST paths. 
Bui et al. \cite{nghi_treecap_2021} proposed TreeCap, a method that uses capsule networks and TBCNNs to learn code models from ASTs.

Graph-based code embedding is another possible approach to capture the semantic and execution aspects of code. 
Fang et al. \cite{fangFunctionalCodeClone2020} developed a joint code representation that combines AST embeddings with CFG and DFG embeddings, using various fusion methods such as concatenation, attention, and gated fusion. 
Guo et al. \cite{graphcodebert2021} introduced GraphCodeBERT, which integrates the data flow of the program into the model, enabling it to learn from both the lexical and the syntactic information of the code. 
Ma et al. \cite{ma_2022_ml} presented a novel model called cFlow, which uses a flow-based Gate Recurrent Unit (GRU) for feature learning from the source-code CFG:
The model leverages the program structure and the semantics of statements along the execution path, which reflects the flowing nature of CFGs.

\subsection{Review of Contrastive Learning}

Contrastive learning is a self-supervised learning technique that leverages the idea of comparing samples against each other to learn meaningful representations of data. 
It aims to learn a vector space where samples that belong to the same class or have similar semantics are close to each other, while samples that belong to different 
classes or have dissimilar semantics are far apart from each other. 
Contrastive learning has been proven effective in various domains where labeled data is scarce or expensive to obtain, such as in computer vision \cite{he2020momentum, chen2020improved,chen2021exploring} and natural language processing \cite{gao-etal-2021-simcse}.

Wu et al. \cite{Wu_2018_CVPR} proposed InstDisc, which used an external memory bank to store negative samples (samples that are different from the query sample). 
However, this approach required a large memory bank and frequent updates, which made it inefficient and impractical. 
For end-to-end learning, InvaSpread \cite{ye2019unsupervised} used only one encoder for computer vision, but its performance was limited by its training batch size, which determined the number of negative samples. 
To address this issue, Oord et al. \cite{oord2018representation} proposed InfoNCE, which generalized contrastive learning to speech and text domains, and introduced a noise-contrastive estimation loss function:
This maximized the mutual information between the query and the positive samples, while minimizing it between the query and the negative samples. 
Tian et al. \cite{tian2020contrastive} proposed CMC, which extended contrastive learning to multi-view or multi-model scenarios --- where the query and the positive samples come from different views or modalities of the same data (such as RGB and depth images, or text and speech).

He et al. \cite{he2020momentum} proposed MoCo v1, which improved on InstDisc \cite{Wu_2018_CVPR} by replacing the external memory bank with a dictionary queue (which stored the encoded representations of previous samples). 
They also introduced a momentum encoder, which updated its parameters slowly to maintain consistency with the dictionary queue. 
MoCo v2 \cite{chen2020improved} further improved MoCo v1 \cite{he2020momentum} by adding more data-augmentation techniques for images (such as color jittering and Gaussian blur) and a projection head, which projected the encoded representations to a lower-dimensional space before computing the contrastive loss. 
MoCo \cite{chen2020improved,he2020momentum} achieved state-of-the-art performance on many computer-vision tasks, such as image classification, object detection, and segmentation. 
Chen et al. \cite{contrastive_chen_2020_mlp} proposed SimCLR v1, which improved on InvaSpread by enlarging the batch size using distributed computing; 
it also added more data-augmentation techniques and a projection head. 
SimCLR v2 \cite{NEURIPS2020_chen_simclr} further improved SimCLR v1 \cite{contrastive_chen_2020_mlp} by pretraining on large unlabeled data using contrastive learning; 
then fine-tuning on task-specific datasets using supervised learning; and 
then finally distilling the encoder using unlabeled data and knowledge distillation. 
Grill et al. \cite{grill2020bootstrap} proposed BYOL and Chen et al. \cite{chen2021exploring} proposed SimSiam,  both of which removed the need for negative samples from contrastive models, enabling them to compare training samples by themselves. 
BYOL \cite{grill2020bootstrap} used another projection head to predict the representation of the query sample, and then minimized the distance between the prediction and the representation of the positive sample. 
SimSiam \cite{chen2021exploring} used a stop-gradient operation to prevent the model from collapsing to a trivial solution, where all samples have the same representation. 
Gao et al. \cite{gao-etal-2021-simcse} proposed SimCSE, which used a contrastive loss function based on cosine similarity to learn sentence embeddings that capture semantic similarity and diversity, without using any additional supervision or data augmentation. 

The research listed above represents the second stage in the development of contrastive learning:
A common approach in this field is to use a momentum encoder and a multi-layer projection (MLP)
head to map the input data into a latent space, where the distance between similar inputs is minimized and the distance between dissimilar inputs is maximized. 
This approach was adopted by both MoCo and SimCLR, two important works in this area \cite{he2020momentum,chen2020improved,contrastive_chen_2020_mlp,NEURIPS2020_chen_simclr}, thus demonstrating its effectiveness. 
Another important aspect of contrastive learning is data augmentation:
This can involve various transformations 
---
such as cropping, flipping, color jittering, and Gaussian noise
---
being applied to the input data to create different (additional) views of it. 
Data augmentation is essential for contrastive learning, as it allows the model to learn features that are invariant to the transformations and discriminative among different instances.

\subsection{Contrastive Learning for Code Embedding}

Many code-embedding approaches rely on supervised learning, which requires a large amount of labeled data.
CodeBERT \cite{codebert2020} and GraphCodeBERT \cite{graphcodebert2021} are pre-trained models that learn from a large corpus of code tokens in different programming languages. However, they are not directly applicable to some SE tasks, such as code-clone detection, that need to predict whether two code fragments are functionally similar or not. 
These tasks require fine-tuning the pre-trained models with additional data and objectives. 
Although pre-trained models can capture the semantics and structure of code effectively, they also have limitations such as the high computational cost and the need for task-specific adaptation.
Wang et al. \cite{HELoC2022} proposed HELoC, a method that uses a hierarchical contrastive loss to predict the level of AST nodes and learn the structural relationships among them. 
They leveraged the hierarchy of ASTs as pseudo-labels, and learned the structural features of source code by predicting the node levels and the three types of hierarchical relationships (neighborhood, adjacent hierarchy and non-adjacent hierarchy relations). 
However, this method is not effective for learning code embeddings, as it only captures node-level relationships and not the entire code's semantics:
It still needs some labeled data to fine-tune on new datasets.
A survey by Chen et al. \cite{Chen2019ALS} revealed a strong demand for code-embedding methods that can leverage unlabeled data, as labeled data is scarce and expensive to obtain.

Contrastive learning aims to bring similar samples closer together, and push dissimilar samples further apart in the vector space. 
Unlike token-level self-supervised learning methods, contrastive learning operates on the level of code snippets, which can improve the performance of code-level software engineering tasks, such as code-clone detection and code classification. 
InferCode, proposed by Bui et al. \cite{buiInferCodeSelfSupervisedLearning2021}, is a contrastive learning method for code embedding that works across different programming languages. 
It first splits the code AST into subtrees of varying sizes, then encodes the subtree nodes with an enhanced TBCNN \cite{mouConvolutionalNeuralNetworks2015}. 
It uses a contrastive loss function to align the embeddings of code snippets that have the same functionality.
Jain et al. proposed ContraCode \cite{Jain2021ContrastiveCR}, which uses a contrastive loss to retrieve syntactically different (but functionally equivalent) variants of a program generated by automated source-to-source compilers.

Contrastive learning is a promising technique for learning code embeddings, but it faces several challenges. 
One challenge relates to generating a sufficient number of transformed samples that preserve the semantic equivalence of the original code snippets. 
Another challenge is the question of how to design an effective encoding network that can capture the syntactic and semantic features of the code. 
A third challenge relates to preventing the model-collapse problem, where the embeddings become indistinguishable from each other. In this paper, we propose some novel solutions to address these challenges: 
First, we introduce a code transformer framework that can generate diverse and semantically equivalent code variants using various code transformation techniques. 
Second, we adopt a Transformer encoder, which has shown great success in code text learning tasks. 
We also investigate the impact of different positional encoding schemes on the performance of the encoder. 
Third, we apply a momentum encoder and a projection head, which have been proven to be efficient ways to improve contrastive learning for computer-vision tasks. 
We show that the momentum encoder can prevent the model-collapse problem, and can enhance the quality of code embeddings.
We demonstrate that our proposed methods can significantly improve contrastive learning for code embedding, and achieve state-of-the-art results on several downstream tasks.

\section{Motivation
  \label{sec_motivation}}

Contrastive learning is a powerful technique for learning representations from images.
It involves training on pairs of images that have the same meaning but different appearances. 
For example, an image of a cat and its cropped, rotated, or color-adjusted version are considered positive pairs.
An image of a cat and an image of a dog are considered negative pairs. 
The goal of contrastive learning is to make the representations of positive pairs more similar and the representations of negative pairs more dissimilar, in a latent space. 
We adopt a similar idea for code:
We transform code snippets, without changing their semantics, and use the original and transformed code as positive pairs for contrastive learning. 
For example, a code snippet and its transformation are considered positive pairs, while two code snippets with different functionality are considered negative pairs. 
The goal of contrastive learning for code is to make the representations of positive pairs more invariant and the representations of negative pairs more discriminative, in a latent space.

This is the motivation of our framework, which is language-independent, as it can handle any programming language.
To achieve this, we use ASTs to perform the code transformations. 
Unlike most existing code-embedding models, which are trained at the token level and have high computational cost and poor generalization, our framework learns code embeddings at the entire code-snippet level, which reduces the model size and facilitates adaptation to new datasets.

In contrast to existing methods, our approach has several novel aspects that make it more effective and efficient for code-representation learning.

\begin{enumerate}
        \item[(1)]
              We normalize all variables in the code ASTs to eliminate the influence of variable naming on model learning.
              This enhances the generalizability of the model.

        \item[(2)]
              We adopt a MoCo-based contrastive learning method \cite{he2020momentum,chen2020improved} to learn code embedding from unsupervised data.
              This makes use of a large, consistent dictionary to encode the query and key embeddings.

        \item[(3)]
              We perform data augmentation by transforming the ASTs.
              This can generate diverse, semantically-equivalent code variants for contrastive learning.
              The transformed ASTs can also serve as augmented data for other supervised learning tasks (as in the experiment in Section~\ref{sec_supervised_learn}).

        \item[(4)]
              We can reduce the learning cost of the model by extracting the critical path of the AST, which contains the most important nodes for code functionality.
              This makes it possible to use encoders with fewer parameters, which can increase the inference speed and reduce memory consumption.

        \item[(5)]
               We employ relative-position encoding and an MLP head \cite{chen2020improved} in our framework's encoder, which can enhance the representation power and capture the structural information of the code.
\end{enumerate}

\section{Code Conversion to Abstract Syntax Tree
  \label{sec_code_to_ast}}

Code differs from natural language in that it has a rigid and precise syntax, with rules that must be obeyed.
Treating code as if it were a natural language would require significantly more effort to learn the structure and semantics.
Data flow analysis and control flow analysis are useful for extracting the structure and semantics of code, but they are usually specific to a certain programming language or SE task.
We, therefore, choose to preprocess the code by converting it into ASTs, preserving both the syntactic structure and the semantic content.
We used tree-sitter\footnote{The \textbf{tree-sitter} is available at: \url{https://tree-sitter.github.io/}.}, which can incrementally parse source files and generate ASTs, even when there are syntax errors in the code.
Because tree-sitter supports multiple programming languages, our framework is also language-independent.

\subsection{Code Normalization}
\label{sec_code_normalsize}

Data normalization is a common technique in deep learning that aims to reduce the influence of data distribution on model performance.
For example, image normalization can help convolutional neural networks to learn features more effectively.
Similarly, code normalization can also improve the quality of code embeddings by eliminating the impact of variable names (which do not impact the code semantics). 
Code normalization can thus also help to reduce the dictionary size of our framework's encoder and make our framework focus more on code structures.
In this paper, we propose a two-step code-normalization method that consists of the following operations:
\begin{itemize}
        \item
              Remove all comments in the code (both line and block comments):
              They do not affect the code execution or functionality.

        \item
              Rename all variables in the code to a standard format that starts with var and ends with a number (var1, var2, var3, \ldots)
              This ensures that the code embeddings are not biased by different naming conventions or styles.
\end{itemize}

\begin{figure}[!ht]
        \centering
        \graphicspath{{img/}}
        \subfigure[Before Normalization]{
                \includegraphics[width=0.225\textwidth]{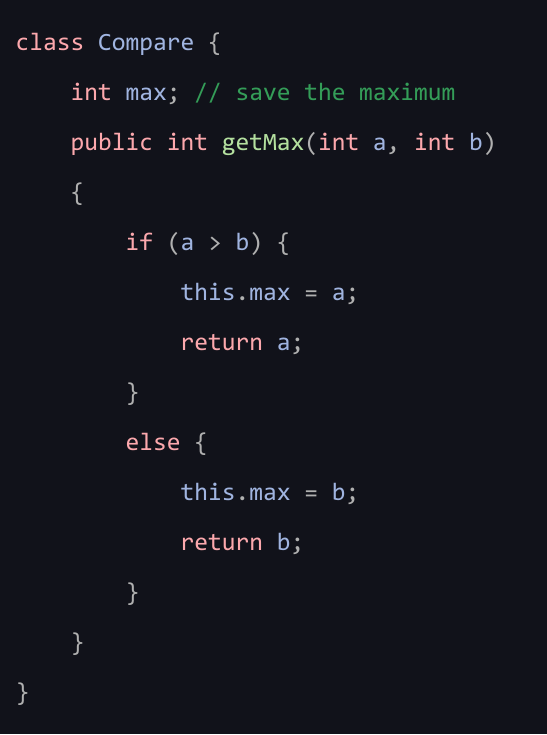}
        }
        \subfigure[After Normalization]{
                \includegraphics[width=0.225\textwidth]{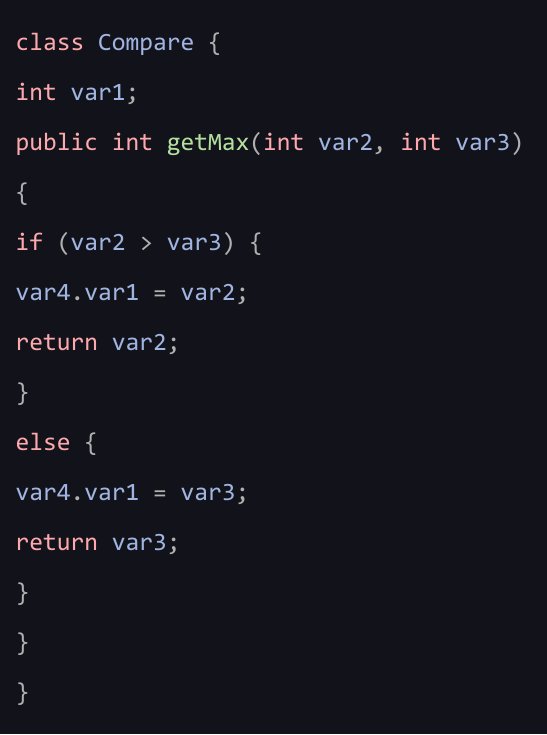}
        }
        \caption{An example of code normalization.}
        \label{fig_normalized}
\end{figure}

Fig.\ref{fig_normalized} shows a comparison of code before and after normalization.

\subsection{Abstract Syntax Tree Transformation}
\label{sec_ast_transformation}

\begin{figure}[!b]
        \centering
        \graphicspath{{img/}}
        \subfigure[Before PermuteDeclaration]{
                \includegraphics[width=0.225\textwidth]{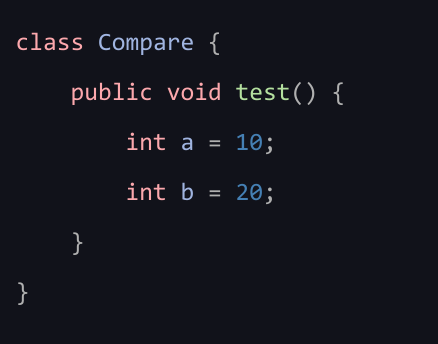}
        }
        \subfigure[After PermuteDeclaration]{
                \includegraphics[width=0.225\textwidth]{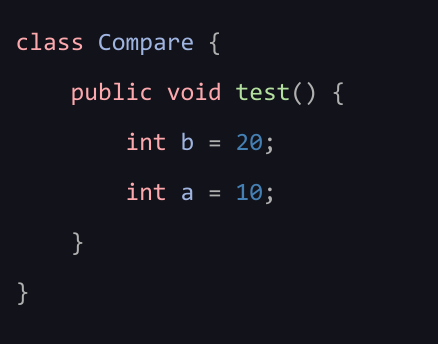}
        }
        \caption{An example of {PermuteDeclaration}.}
        \label{fig_permute_declare}
\end{figure}

\begin{figure}[!b]
        \centering
        \graphicspath{{img/}}
        \subfigure[Before SwapCondition]{
                \includegraphics[width=0.225\textwidth]{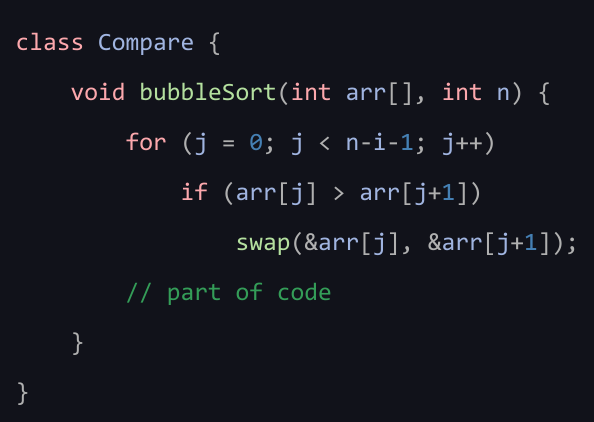}
        }
        \subfigure[After SwapCondition]{
                \includegraphics[width=0.225\textwidth]{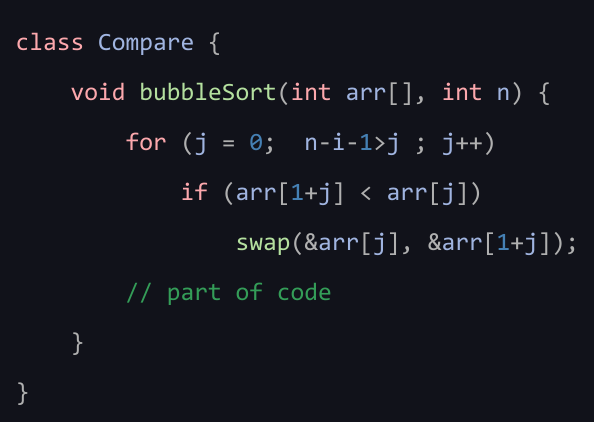}
        }
        \caption{An example of {SwapCondition}.}
        \label{fig_swap}
\end{figure}

\begin{figure}[!t]
        \centering
        \graphicspath{{img/}}
        \subfigure[Before ArithmeticTransform]{
                \includegraphics[width=0.225\textwidth]{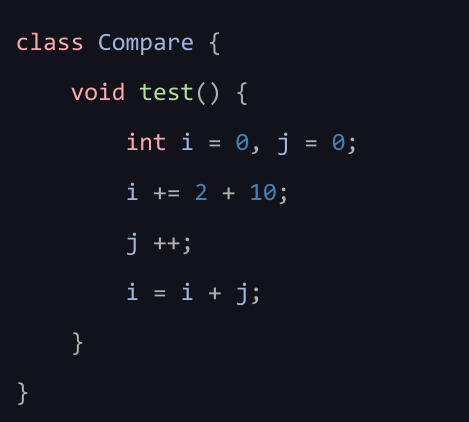}
        }
        \subfigure[After ArithmeticTransform]{
                \includegraphics[width=0.225\textwidth]{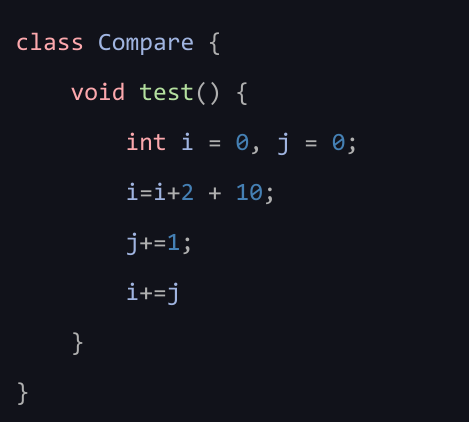}
        }
        \caption{An example of {ArithmeticTransform}.}
        \label{fig_algorithm}
\end{figure}

\begin{figure}[!b]
        \centering
        \graphicspath{{img/}}
        \subfigure[Before While statement]{
                \includegraphics[width=0.225\textwidth]{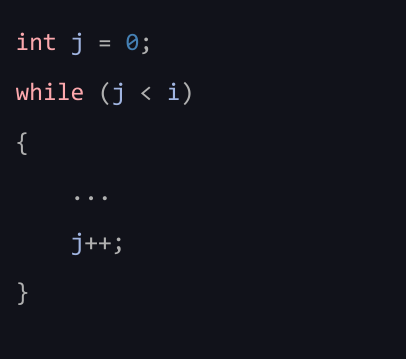}
        }
        \subfigure[After For statement]{
                \includegraphics[width=0.225\textwidth]{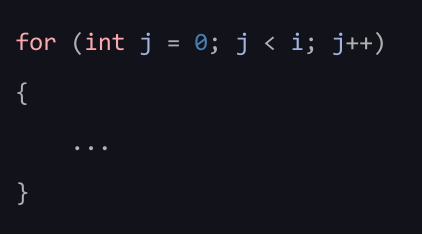}
        }
        \caption{An example of {WhileForExchange}.}
        \label{fig_while_for}
\end{figure}

Unsupervised representation learning is a challenging task that aims to learn meaningful and generalizable features from unlabeled data.
Contrastive learning is a popular technique for this task, especially in computer vision, where it leverages data augmentation to create positive and negative pairs of samples.
The idea is to train a model that can learn embeddings that are close together for positive pairs (similar samples) and far apart for negative pairs (dissimilar samples).
In this paper, we extend this idea to code-representation learning, where we use ASTs as the input.
ASTs are tree structures that capture the syntactic and semantic information of the code.
We perform data augmentation on the ASTs by applying transformations, such as swapping, inserting, or deleting nodes, to create new ASTs, called \textbf{anchor ASTs}.
These anchors are similar to the original ASTs in terms of functionality, but different in terms of syntax.
Our framework learns 
the parameters for code embeddings using contrastive learning
---
the original and anchor ASTs are close together in the embedding space, but different ASTs are far apart.
We show (in Section \ref{sec_visualization})
that our method can learn robust and discriminative code embeddings that outperform existing methods on several code analysis tasks.

\begin{figure}[!ht]
        \centering
        \graphicspath{{img/}}
        \subfigure[Before AddDummyStatement]{
                \includegraphics[width=0.225\textwidth]{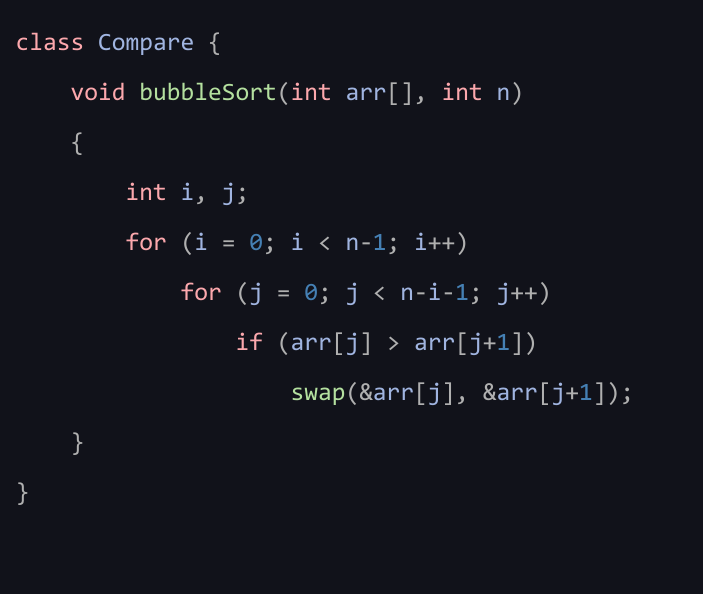}
        }
        \subfigure[After AddDummyStatement]{
                \includegraphics[width=0.225\textwidth]{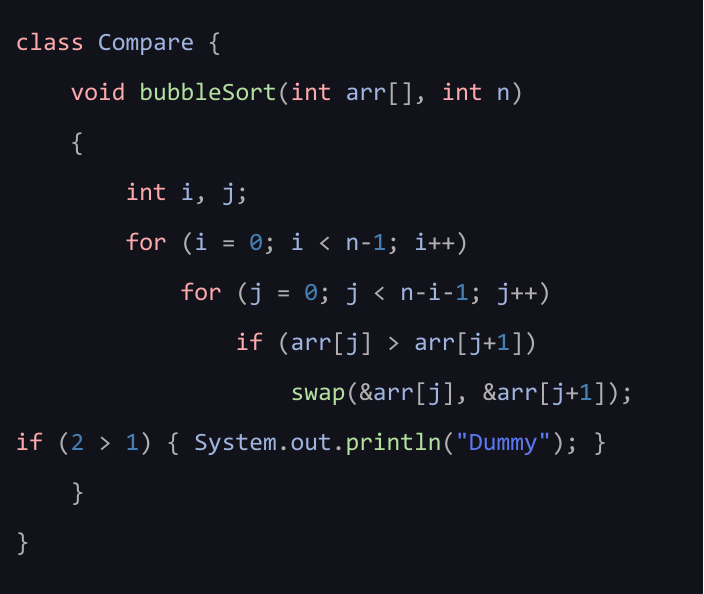}
        }
        \caption{An example of {AddDummyStatement}.}
        \label{fig_dummy}
\end{figure}

\begin{figure}[!t]
        \centering
        \graphicspath{{img/}}
        \subfigure[Before AddTryCatch]{
                \includegraphics[width=0.225\textwidth]{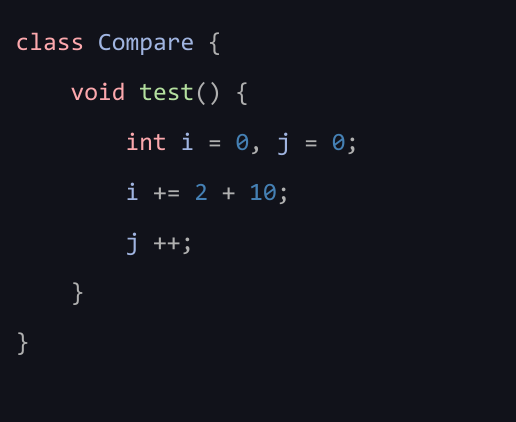}
        }
        \subfigure[After AddTryCatch]{
                \includegraphics[width=0.225\textwidth]{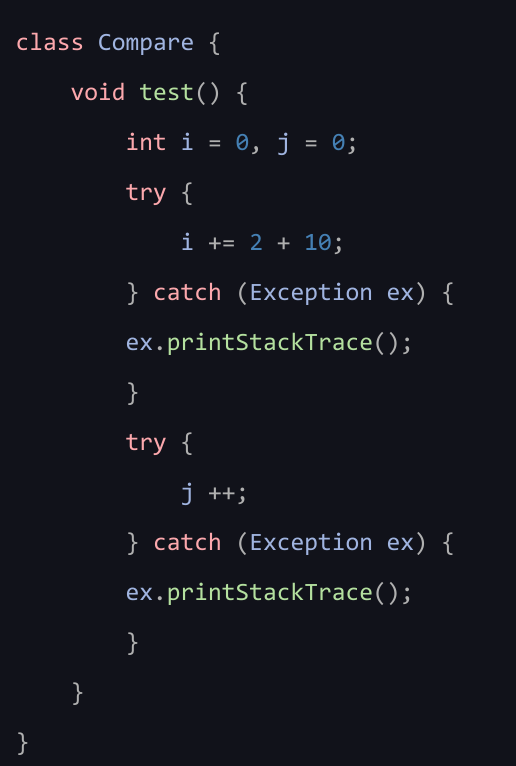}
        }
        \caption{An example of {AddTryCatch}.}
        \label{fig_try_catch}
\end{figure}

Producing anchor ASTs involves modifying the code syntax without changing its semantics.
This transformation can be used to create diverse and realistic code samples for contrastive learning.
However, different programming languages may have different syntax rules and features, which may require different transformation methods.
In this paper, we propose a set of common code transformation methods that can be applied to most programming languages, while preserving their semantics.
Although some specific programming languages may have certain semantics-preserving transformation methods that are unique to their syntax, in this paper, we focus on the common transformation methods that are sufficient for our contrastive learning framework.
We obtain the anchor ASTs through a combination of the following transformation methods:
\begin{enumerate}
        \item
              \textbf{PermuteDeclaration}, as shown in Fig.~\ref{fig_permute_declare}, switches the order of variable declaration statements.

        \item
              \textbf{SwapCondition}, as shown in Fig.~\ref{fig_swap}, swaps the two operands of the binary operator, keeping the semantics unchanged, like $a > b$ becomes $b < a$ and $j+1$ changes into $1+j$.

        \item
              \textbf{ArithmeticTransform}, as shown in Fig.~\ref{fig_algorithm}, converts arithmetic operations into different forms.

        \item
              \textbf{WhileForExchange}, as shown in Fig.~\ref{fig_while_for}, converts \textit{while} statements into \textit{for} statements.

        \item
              \textbf{AddDummyStatement}, as shown in Fig.~\ref{fig_dummy}, adds dead code to a random selection of statements, but does not change the code semantics.

        \item
              \textbf{AddTryCatch}, as shown in Fig.~\ref{fig_try_catch}, adds \textit{try-catch} statements to a random selection of statements, but does not change the code semantics.

        \item
              \textbf{PermuteStatement} is similar to \textbf{PermuteDeclaration}, but exchanges randomly selected statements.
              Only a small percentage of statements are selected.
\end{enumerate}

\subsection{Abstract Syntax Tree Extraction}
\label{sec_code_extract}

To identify the critical execution paths from the original and anchor ASTs, we use a set of rules that traverse the ASTs in a depth-first manner, and prune nodes that do not contribute to the syntactic or semantic meaning of the code.
This process results in a simplified and streamlined AST that captures the essential logic of the code.
We then obtain a sequence of tokens by concatenating them in the order of their appearance in the depth-first traversal of the simplified AST.

\begin{figure}[!t]
        \begin{lstlisting}[language=Java]
public class BubbleSortExample {
    static void bubbleSort(int[] arr) {  
        int n = arr.length;  
        int temp = 0;  
        for(int i=0; i < n; i++) {  
            for(int j=1; j < (n-i); j++) {
                if(arr[j-1] > arr[j]){
                    // swap elements  
                    temp = arr[j-1];  
                    arr[j-1] = arr[j];  
                    arr[j] = temp;  
                }            
            } 
        }
    }
}
\end{lstlisting}
        \caption{A Java implementation of the Bubble Sort algorithm.}
        \label{fig_example_program}
\end{figure}

Fig.~\ref{fig_example_program} shows a Java implementation of the Bubble Sort algorithm \cite{sorting_1956}.
Extraction of the AST tokens from the Bubble Sort program can be done as described in the previous paragraph.
The expressions and statements in Fig.~\ref{fig_example_program} can be simplified by removing unnecessary parentheses, etc.
Fig.~\ref{fig_simplified} shows the simplified AST tokens obtained from the Java program in  Fig.~\ref{fig_example_program}:
The tokens capture the main logic of the sorting procedure, such as the nested loops, the comparison and swapping of array elements, and the assignment of variables.

\begin{figure}[!t]
        \begin{lstlisting}
    'n', '=', 'arr.length', 'temp', '=', '0', 'for', 'i', '<', 'n', 'i', '++', 'i', '=', '0', 'for', 'j', '<', 'j', '++', 'j', '=', '1', 'if', 'n', '-', 'i', 'arr[j-1]', '>', 'arr[j]', 'temp', '=', 'arr[j-1]', 'arr[j-1]', '=', 'arr[j]', 'arr[j]', '=', 'temp'
        \end{lstlisting}
        \caption{Simplified AST tokens after extraction without normalization.}
        \label{fig_simplified}
\end{figure}

To make the model more robust and generalizable, a normalization step is applied before extracting the AST tokens.
The normalization step replaces all variable names with generic placeholders (``var1'', ``var2'', etc.).
This allows the model to focus on the syntactic and semantic features of the code, rather than the specific naming choices of the programmer.
Normalization takes place in both the training and inference phases, before the AST extraction.
Fig.~\ref{fig_simplified_normalized} shows an example of the simplified AST tokens after normalization.
The tokens do not require any special tokens to indicate the start or end of the sequence and can be directly fed into the model as input.

\begin{figure}[!t]
        \begin{lstlisting}
    'var2', '=', 'var1.var5', 'var3', '=', '0', 'for', 'var4', '<', 'var2', 'var4', '++', 'var4', '=', '0', 'for', 'var6', '<', 'var6', '++', 'var6', '=', '1', 'if', 'var2', '-', 'var4', 'var1[var6-1]', '>', 'var1[var6]', 'var3', '=', 'var1[var6-1]', 'var1[var6-1]', '=', 'var1[var6]', 'var1[var6]', '=', 'var3'
    \end{lstlisting}
        \caption{Simplified AST tokens after extraction with normalization.}
        \label{fig_simplified_normalized}
\end{figure}

\subsection{Sample Generation}
\label{sec_sample_generation}

Initially, each training sample, denoted as \( Code_{train} \), undergoes a process of code normalization (Section~\ref{sec_code_normalsize}), resulting in a normalized code set \( Code_{normalized} \). Subsequently, the anchor sample \( Code_{anchor} \) is derived from \( Code_{normalized} \) by applying Abstract Syntax Tree (AST) transformations (Section~\ref{sec_ast_transformation}). Positive sample pairs are then generated by extracting the critical execution paths within the ASTs of both \( Code_{normalized} \) and \( Code_{anchor} \) (Section~\ref{sec_code_extract}).

Throughout the training phase, negative samples are identified as any sample that is not the positive sample (anchor sample) for the current training iteration. This approach ensures a clear distinction between positive and negative samples, facilitating the model's learning process. Following each training epoch, negative samples are entered into a queue, as explained in Section~\ref{sec_sub_framework}.

\section{TransformCode
  \label{sec_framework}}

In this paper, we propose TransformCode, a novel unsupervised learning framework for code embedding based on contrastive learning.
Contrastive learning is a branch of self-supervised learning that does not require any human-annotated labels or supervision:
Models can generate their own labels or pseudo-labels from the data itself, and use them to learn useful representations or features. 
Our proposed framework can also produce transformed codes as pseudo-labels for training.

\begin{figure*}[!t]
        \centering
        \graphicspath{{img/}}
        \includegraphics[width=0.95\textwidth]{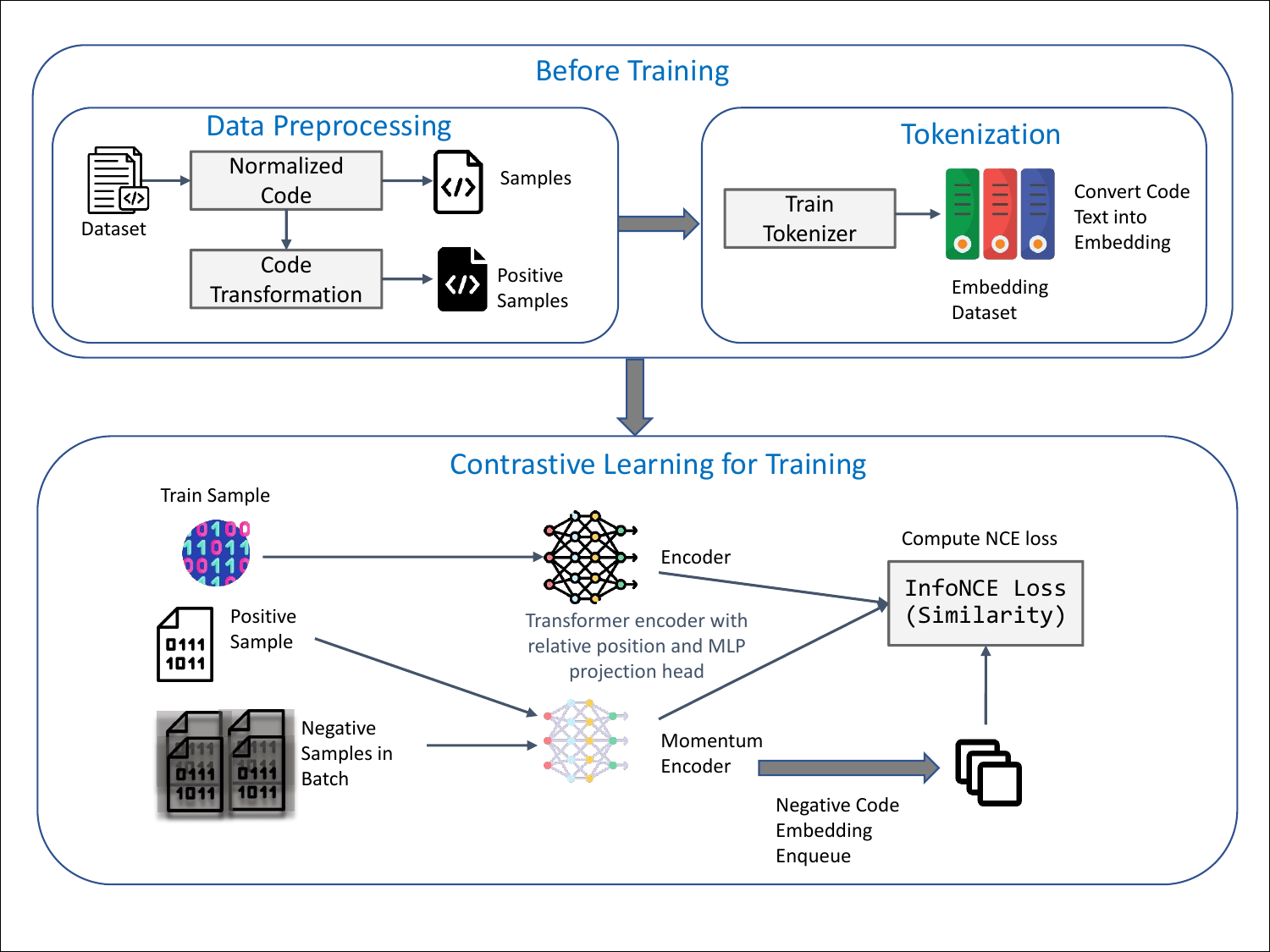}
        \caption{Our proposed framework for unsupervised learning of code embedding.}
        \label{fig_proposed}
\end{figure*}

\begin{figure*}[!b]
        \centering
        \graphicspath{{img/}}
        \subfigure[CodeBERT for original code]{
                \includegraphics[width=0.315\textwidth]{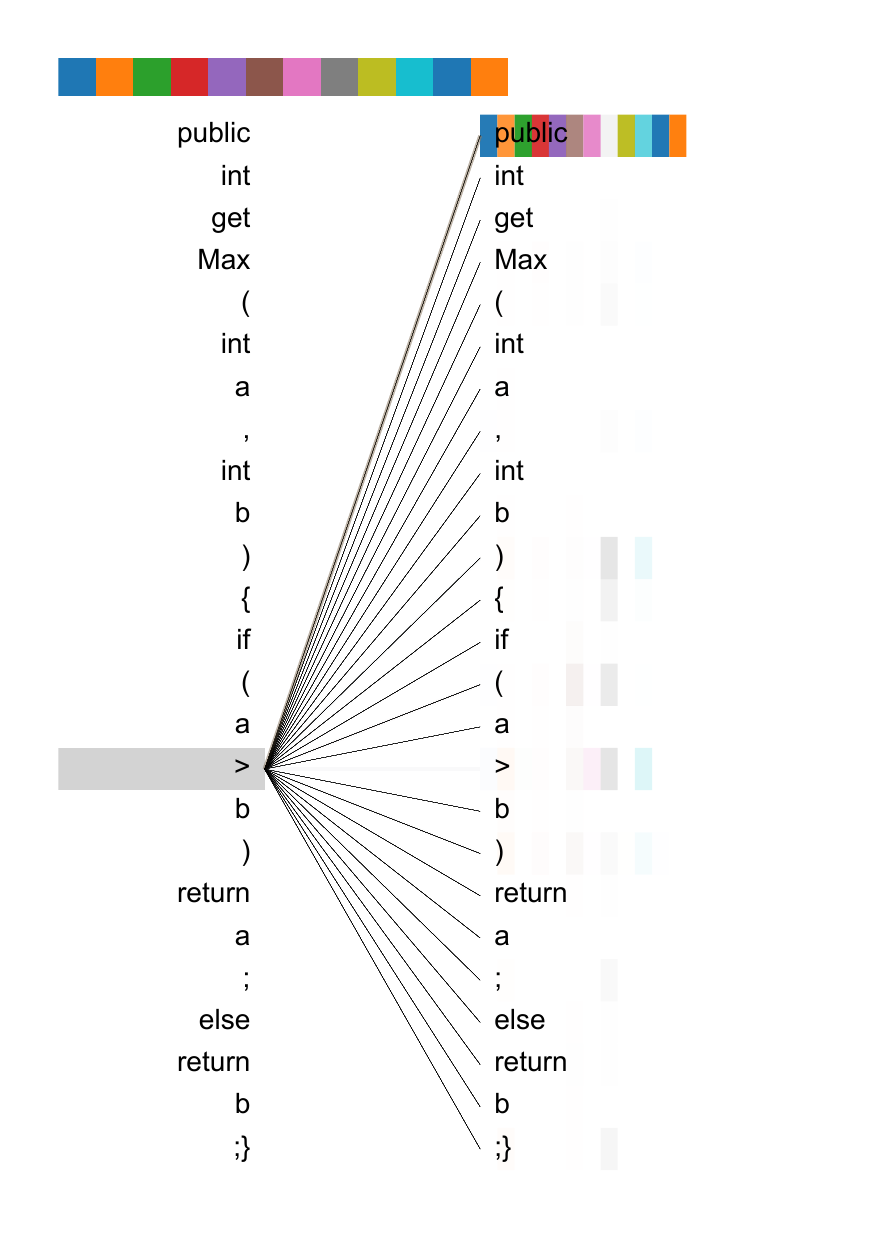}
        }
        \subfigure[CodeBERT for original code with begin and end of sentence tokens]{
                \includegraphics[width=0.315\textwidth]{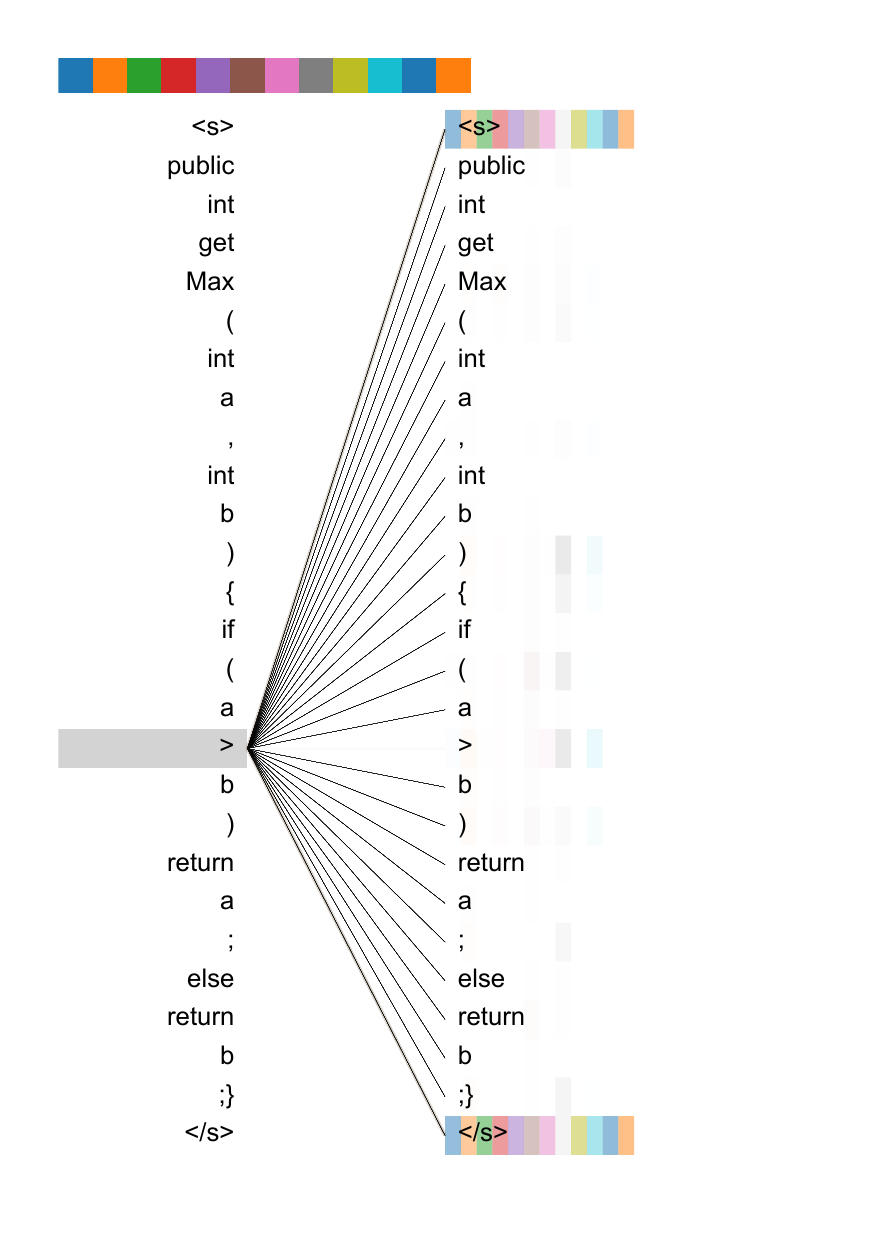}
        }
        \subfigure[CodeBERT with begin and end of sentence tokens]{
                \includegraphics[width=0.315\textwidth]{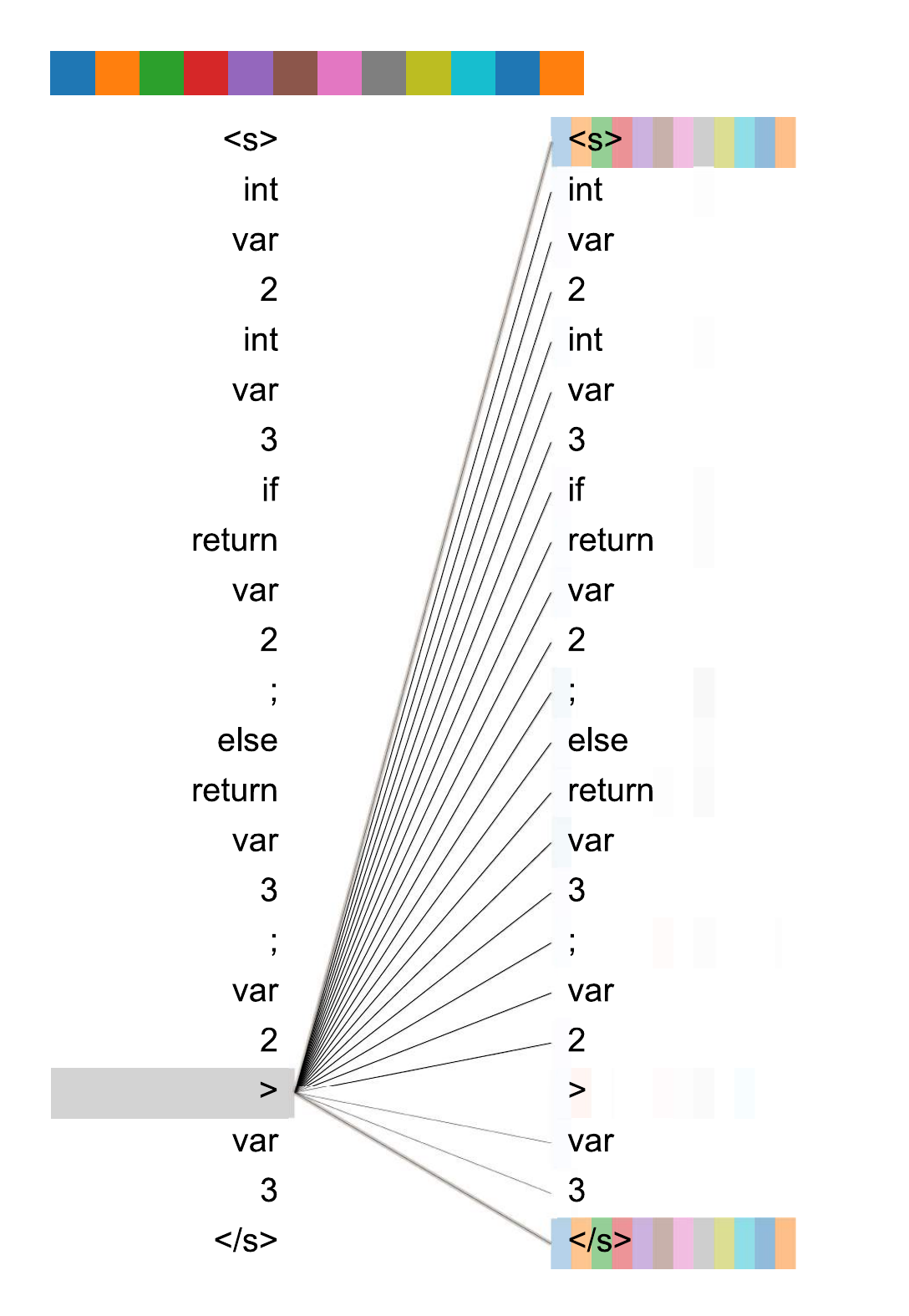}
        }
        \subfigure[CodeBERT without begin and end of sentence tokens]{
                \includegraphics[width=0.315\textwidth]{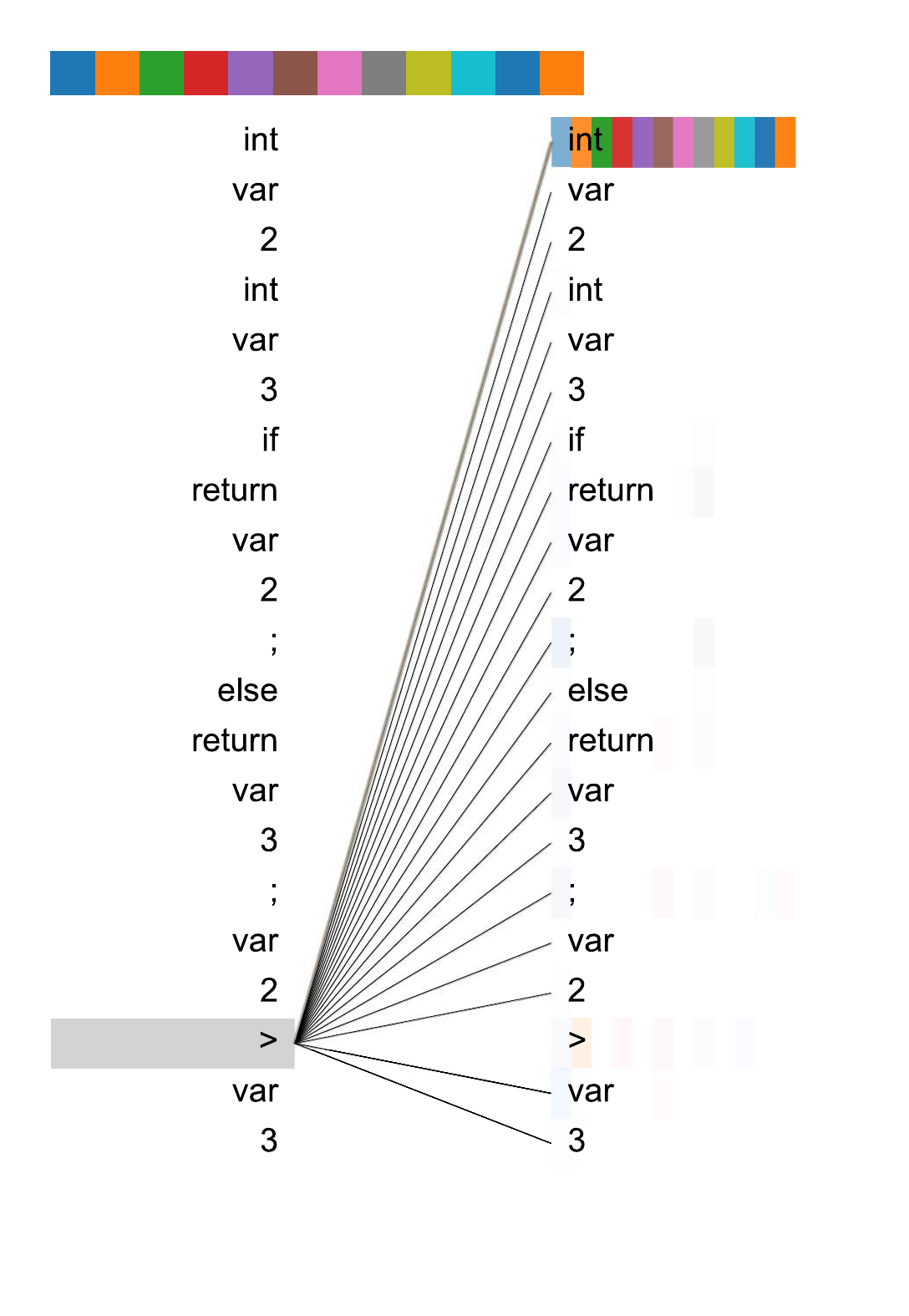}
        }
        \subfigure[Our proposed method]{
                \includegraphics[width=0.315\textwidth]{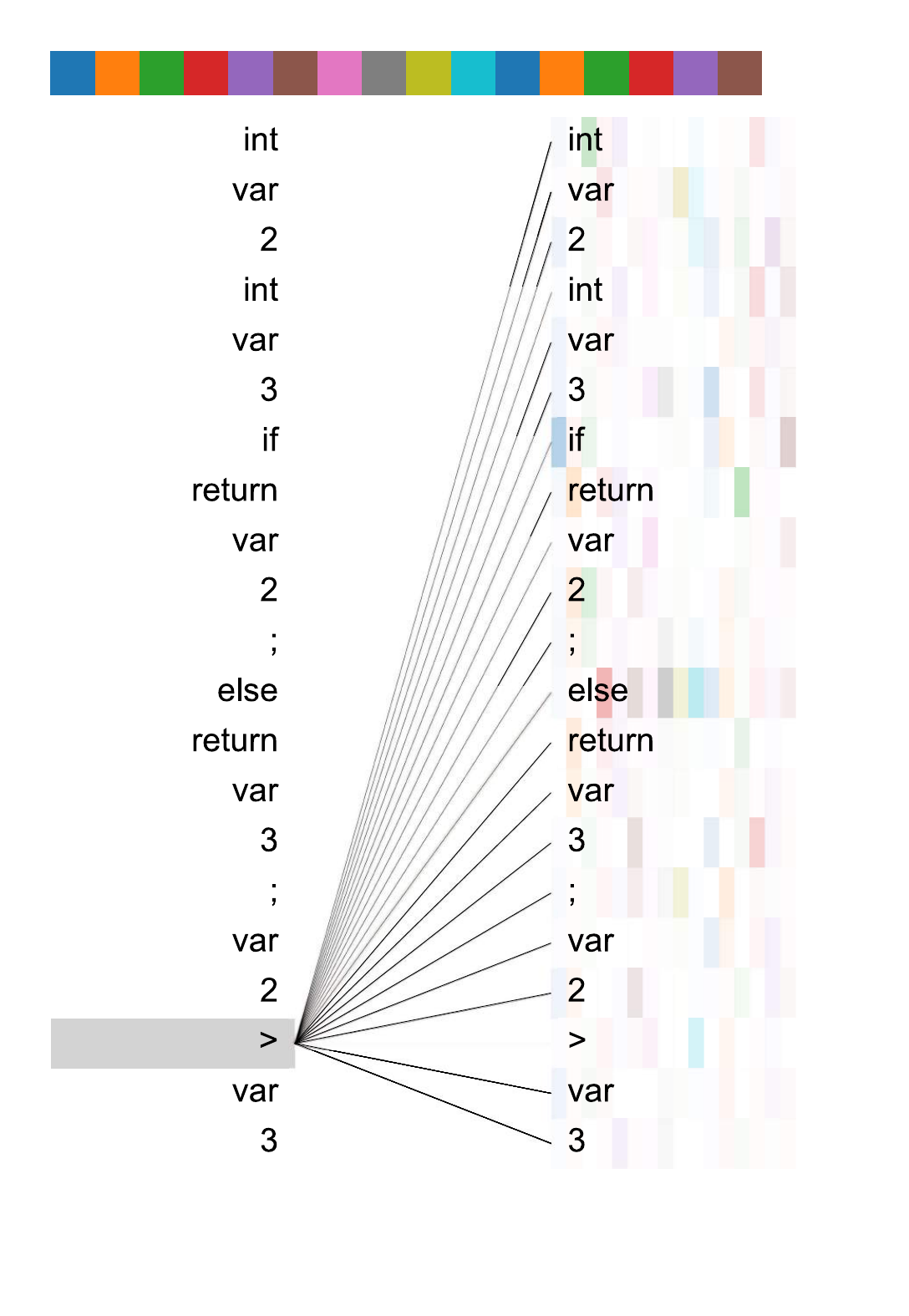}
        }
        \caption{Comparisons between our proposed method and fine-tuning pre-trained {CodeBERT} using Bertviz \cite{vig-2019-multiscale}.}
        \label{fig_codebert}
\end{figure*}

\subsection{Framework}
\label{sec_sub_framework}

Our framework comprises three main steps, as shown in Fig.~\ref{fig_proposed}. 
The first step involves preprocessing code snippets to obtain the normalized and transformed code, which is used as the training data. 
The second step is to train the code tokenizer (as explained in Section \ref{sec_tokenizer}), and to embed all code snippets using the trained tokenizer. 
The third step is training models, which has both a query encoder and a momentum encoder --- the encoders have the same architecture, but different parameters. 
The momentum encoder is updated by a moving average of the query encoder, which helps to maintain the consistency and diversity of the representations. This also helps to avoid the collapse of the contrastive loss. Our framework is encoder-agnostic, allowing for the use of any encoder. 
In this paper, we employ the Transformer encoder \cite{vaswani2017attention} as our encoder, due to its powerful performance in natural language processing.

We first build a queue of negative samples for each query. 
The queue is maintained as a first-in-first-out (FIFO) buffer, where the encoded representations of the current mini-batch are enqueued, and the oldest ones are dequeued.
Each training sample is treated as a query to find the best matching key from each input batch and the queue of code snippets. 
The best matching key is obtained by applying code transformations to the current code sample, which forms the positive pair (also known as the anchor pair in contrastive learning). 
The best matching key is also called the positive sample (anchor sample).
We also form negative pairs by pairing the query sample with the other keys:
The negative samples are all other code samples in the mini-batch and in the queue.
We use the InfoNCE loss \cite{oord2018representation} to maximize agreement between the positive pair, and minimize the agreement between the negative pairs.

InfoNCE loss is defined Eq.~(\ref{eq_infonce}):
\begin{equation}
        \small
        \mathcal{L}_{q, k^{+},\left\{k^{-}\right\}}=-\log \frac{\exp \left(q \cdot k^{+} / \tau\right)}{\exp \left(q \cdot k^{+} / \tau\right)+\sum_{k^{-}} \exp \left(q \cdot k^{-} / \tau\right)}
        ,
        \label{eq_infonce}
\end{equation}
where $q$ is a query code representation (training sample);
$k^{+}$ is a code representation of the positive key sample (anchor sample);
$k^{-}$ are representations of the negative key samples (dissimilar samples);
and $\tau$ is a temperature parameter that controls the softness of the softmax function.
InfoNCE loss aims to maximize the likelihood of the positive pair relative to the negative pairs, which is equivalent to maximizing the lower bound of the mutual information between $q$ and $k^+$. 
(For more details, the interested reader is referred to the work of van den Oord et al.~\cite{oord2018representation}.)
It should be noted that the query sample should be normalized, and the positive key and negative key samples should be normalized and transformed.
By minimizing the InfoNCE loss, the query and anchor samples are brought closer together in the learning representation space, and the negative keys are pushed further apart.

\subsection{Model Architecture}

We designed our framework to be encoder-agnostic, which makes it possible to use any encoder.
In this paper, we adopt the transformer encoder as our AST encoder $E$ \cite{vaswani2017attention}, 
but replace the sinusoidal position encoding with relative-position encoding \cite{shaw-etal-2018-self}.
The AST encoder is a Transformer-based model that encodes both the original and the transformed code snippets into a latent space. 
The encoder consists of an embedding layer, a relative-position encoding layer, and a stack of N encoder layers. 
The embedding layer maps each token in the code snippet to a d-dimensional vector. 
The relative-position encoding layer adds a relative-position vector to each token embedding: This captures the syntactic structure and the relative distance between tokens in the code snippet. 
The encoder layer is composed of a multi-head self-attention sublayer and a feed-forward sublayer, followed by layer normalization and a residual connection. 
The multi-head self-attention sublayer allows the model to attend to different parts of the code snippet and learn the semantic relationships between tokens. 
The feed-forward sublayer consists of two linear transformations with a ReLU activation in between:
This enhances the model's expressive ability and non-linearity.
The relative position encoding is defined in Eq.~(\ref{eq_rpe}).
\begin{equation}
        \small
        e_{i j}=\frac{x_i W^Q\left(x_j W^K\right)^T+x_i W^Q\left(a_{i j}^K\right)^T}{\sqrt{d_z}},
        \label{eq_rpe}
\end{equation}
where $a_{ij}$ is the edge between the input elements $x_{i}$ and $x_{j}$.
More details can be found in the work by Shaw et al.~\cite{shaw-etal-2018-self}.

Position encoding enables the model to distinguish between identical words in different positions and to produce position-aware representations. This also helps the model to better capture the relative position structure of the code. A simplified AST is used (Section~\ref{sec_code_extract}), which enables our framework to capture more syntactic and semantic information about the code. To further capture the code's structural information and improve the contrastive learning performance, we introduce an MLP head $P$ \cite{chen2020improved, contrastive_chen_2020_mlp} after the encoder $E$. The MLP head is a stack of M linear layers with a ReLU activation after each layer except the last one. The MLP head projects the output of the encoder to a lower-dimensional space $h_i$, where the representations of positive pairs are more invariant and the representations of negative pairs are more discriminative. 
The MLP head can be formulated as shown in Eq.~(\ref{eq_mlp}).
\begin{equation}
\begin{aligned}
    h_0 &= E (x) \\
    h_i &= \sigma(W_i h_{i-1} + b_i), \quad i = 1, \dots, M-1 \\
    z &= W_M h_{M-1} + b_M
\label{eq_mlp}
\end{aligned}
\end{equation}
where $x$ is the input; $E$ is the encoder; $W_i$ and $b_i$ are the weight matrix and bias vector of the $i$-th linear layer; $h_i$ is the output of the $i$-th linear layer; and $z$ is the final code representation. The MLP head can be seen as a non-linear transformation that enhances the code representation's quality and robustness. Moreover, the MLP head also interacts with the relative-position encoding, as the relative-position vector is added to each token embedding before being fed into the code snippet encoder. The relative-position encoding can provide useful structural information to the MLP head.
In our proposed model, we train a base encoder network $E$ and a projection head $P$ to maximize agreement using a contrastive loss. 
We discuss the effects of using position encoding and MLP heads in the ablation experiment in Section~\ref{sec_exp_unsupervised}.

\subsection{Limitations of Pre-trained Models}

Recent advances in open-source pre-trained models (PTMs) and parameter-efficient fine-tuning (PEFT) methods, such as LoRA \cite{hu2022lora}, have made it possible to fine-tune PTMs for downstream tasks.
However, these methods still face some limitations when applied to code embedding.
For instance, most PTMs like CodeBERT use special tokens, such as $<s>$ and $</s>$, to mark the start and end of sentences.
This causes the model to pay more attention to these tokens than to the actual code content, even after extracting the streamlined AST paths using our method.
As we will show in Section~\ref{sec_visualization}, this results in a suboptimal performance of the fine-tuned CodeBERT model for 
code-clone detection.
To overcome this issue, we propose a novel method that does not require any pre-trained model or special tokens, but trains a new model from scratch using our extracted AST paths.
The Java code snippet in Fig.~\ref{fig_sample_java_code} demonstrates our approach: 
The upper part of the figure shows the original code, and the normalized code with the execution path extracted is shown in the lower part.
\begin{figure}[!t]
    \begin{lstlisting}[language=Java]
public int getMax(int a, int b) { if (a>b) return a; else return b;}
    \end{lstlisting}
    \begin{lstlisting}[language=Java]
int var2 int var3 if return var2; else return var3; var2 > var3
    \end{lstlisting}
    \caption{A sample of Java code.}
    \label{fig_sample_java_code}
\end{figure}

Our method can capture the semantic structure and logic of the code more effectively, as illustrated by the attention weights in Fig.~\ref{fig_codebert}
---
the different colors in the figure represent different attention heads (the model has 12), and
the darker shades indicate higher attention weights.
Fig.~\ref{fig_codebert} shows that our method allocates attention more uniformly, and selectively attends to the crucial parts of the code, such as the conditional branches (if-else branches).
CodeBERT, in contrast, tends to concentrate more on the sentence beginnings, irrespective of the input format, or the presence of start and end tokens. 
This indicates that our approach is more efficient than CodeBERT, because our method can use fewer resources and achieve better performance than CodeBERT.

We will demonstrate, in Section \ref{sec_exp_unsupervised},
that our method has several advantages over the existing pre-trained code-representation model, CodeBERT.
Firstly, our method converges faster and achieves better results than the fine-tuned CodeBERT model for code embedding.
Secondly, our method uses a smaller model architecture than CodeBERT, which reduces the number of parameters and the computational costs:
Specifically, our proposed model has either two or four layers with 1024 or 768 hidden sizes,
while CodeBERT has 12 layers with 768 hidden sizes.
This means that training our proposed model requires less GPU memory and power consumption.
CodeBERT was trained on 8 NVIDIA V100 GPUs for 1 million steps, with a batch size of 256 and a learning rate of 1e-4. The training for CodeBERT took about 4 days from scratch \cite{codebert2020}. Our proposed model is smaller than CodeBERT, and can be trained from scratch in several hours on 2 NVIDIA 3090 GPUs. 
Thirdly, our method adopts a different training objective than the traditional pre-trained language model:
We do not use masked language models (MLMs) or next sentence prediction (NSP) to learn token-level representations, which may not capture the global semantics of the code.
Instead, it is more effective when contrastive learning is applied to the whole representation of the code snippets:
This can better preserve the functionality and structure of the code.
We believe that our method is a novel and effective way to learn code vector spaces from unlabeled data.

\subsection{Tokenizer}
\label{sec_tokenizer}

Tokenization is an essential step in natural language processing, as it allows the model to break down the input text into smaller units that can be processed more efficiently.
For code embedding, which aims to learn meaningful representations of the source code, tokenization is also a crucial step, as it determines how the code is split into tokens that capture its syntactic and semantic features.
There are three main types of tokenization methods:
word-based;
character-based; and
subword-based.

Word-based tokenization relies on splitting the text by whitespace and punctuation, which may result in a large vocabulary size due to the diversity and complexity of natural language and code.
Furthermore, word-based tokenization cannot handle out-of-vocabulary (OOV) words, or learn the relationships between different words.

Character-based tokenization treats each character as a token, thus avoiding the problem of OOV words.
However, this also leads to a very long sequence of tokens that may be difficult for the model to learn from.
Character-based tokenization may also lose some information at the word level, such as morphology and semantics.
Therefore, subword-based tokenization, because it combines the advantages of word-based and character-based tokenization, is the most popular choice in current pre-training models

Subword-based tokenization splits the text into smaller units (morphemes) that are frequent and meaningful, such as prefixes, suffixes, and stems.
Some common subword-based tokenization algorithms are:
Byte-Pair Encoding (BPE) \cite{sennrich-etal-2016-neural};
WordPiece \cite{wordpiece,devlin2018bert}; and
Unigram \cite{kudo-2018-subword}.
CodeBERT \cite{codebert2020}, a pre-trained model for code understanding and generation, adopts the same model architecture and tokenization method as RoBERTa \cite{liu2019roberta}, which is a state-of-the-art natural language understanding model based on BERT \cite{devlin2018bert}.
BPE is a data-driven algorithm that iteratively merges the most frequent pairs of characters or subwords until a predefined vocabulary size is reached.
However, RoBERTa's BPE vocabulary contains many non-ASCII characters that are not related to the code, which may negatively impact on the model's performance and efficiency.
We, therefore,  use WordPiece to train our own tokenizer on a large-scale code dataset.
WordPiece thus considers the linguistic information of the data, and may produce more meaningful subwords.
We hypothesize that WordPiece tokenization may be more suitable for tasks that require code embedding, as it can better capture the syntax and semantics of the code.
To verify our hypothesis, we conducted experiments on several code-related tasks and compared our WordPiece tokenizer with the original CodeBERT tokenizer.
We found that our WordPiece vocabulary had less than 20,000 subwords, which is far fewer than CodeBERT's vocabulary of more than 50,000.
This reduces the computational cost and memory usage of the model.
We also observed that our framework converged faster, and achieved more stable results than CodeBERT on most tasks, which demonstrates the effectiveness of our WordPiece tokenizer for code embedding.

\section{Experiments
  \label{sec_experiment}}

\subsection{Evaluation Metrics and Environment}

Code-clone detection is a classification task that involves determining whether or not two code fragments are the same.
To evaluate the performance of code-clone detection, we use the following metrics that are commonly used in classification tasks
($TP$ represents true positives;
$TN$, true negatives;
$FP$, false positives; and
$FN$ represents false negatives):

\begin{equation}
        \small
        Accuracy =  \left(\frac{TP+TN}{TP+TN+FP+FN}\right),
        \label{eq_acc}
\end{equation}

\begin{equation}
        \small
        Precision = \left(\frac{TP}{TP+FP}\right),
        \label{eq_prec}
\end{equation}

\begin{equation}
        \small
        Recall = \left(\frac{TP}{TP+FN}\right),
        \label{eq_recall}
\end{equation}

\begin{equation}
        \small
        F1 = 2 \times \frac{Precision \times Recall}{Precision + Recall}.
        \label{eq_f1}
\end{equation}

Accuracy measures the proportion of correct predictions among all predictions, as defined in Eq.~(\ref{eq_acc});
Precision measures the proportion of positive predictions that are actually positive, as defined in Eq.~(\ref{eq_prec});
Recall measures the proportion of positive instances that are correctly predicted, as defined in Eq.~(\ref{eq_recall}); and
F1 is the harmonic mean of precision and recall, which balances both metrics, as defined in Eq.~(\ref{eq_f1}).

Accuracy may not be a reliable metric when dealing with imbalanced datasets \citep{imbalance_data_2013}, where some classes or categories are underrepresented or overrepresented. In such cases, Accuracy may be biased by the dominant class, resulting in ignoring the minority class.
For example, if a dataset has 90\% positive instances and 10\% negative instances, a classifier that always predicts positive will have 90\% accuracy, but it will fail to detect any negative instances.
Therefore, Accuracy may not reflect the true performance of the classifier on imbalanced datasets.
To address this, we use the F1 Score (the harmonic mean of precision and recall) as an alternative metric for imbalanced datasets \citep{imbalance_data_2013}.
The F1 Score takes into account both precision and recall:
It gives a higher value when both precision and recall are high, which means that the classifier can correctly identify both positive and negative instances.
The F1 Score is lower when either precision or recall is low, which indicates that the classifier either misses some positive instances or produces some false positives.
In summary, the F1 Score provides a more accurate and robust measure of the classifier's performance on imbalanced datasets.

Because the dataset for the prediction experiment was imbalanced (with respect to different method names), we used the F1 Score in the evaluations.
In contrast, because the code classification experiment used a balanced dataset (with respect to different code categories), we were able to use Accuracy as the evaluation metric.

We followed the original studies \cite{buiInferCodeSelfSupervisedLearning2021,zhangNovelNeuralSource2019} to guide the choice of evaluation metrics for these experiments.
We also copy the original experimental results \cite{buiInferCodeSelfSupervisedLearning2021,zhangNovelNeuralSource2019}, for ease of comparison.
All our experiments were conducted on an AMD 5700X computer with two Nvidia RTX 3090 graphics cards.

\subsection{Data Augmentation for Experiments}

All anchor samples were generated using multi-threaded processing. Specifically, on an AMD 5700X CPU, the process utilized 16 threads. Upon completion of the data preprocessing, the data was serialized and stored on disk for future experiments. In the context of the BigCloneBench dataset, which was employed for unsupervised clone detection, data preprocessing was completed in 14 seconds for 7302 samples. Training durations were equally efficient, requiring only 17 seconds per epoch with a batch size of 128. Regarding the OJClone C dataset, data preprocessing was completed in approximately 90 seconds for 52,000 samples. The training phase, on the other hand, required around 3 minutes per epoch, also with a batch size of 128. The method-name prediction experiment necessitated a slightly modified preprocessing approach, as it involved extracting the method name for each method. Consequently, data preprocessing required roughly 7 minutes for 89,393 samples. Training times were observed to be around 11 minutes per epoch when employing a batch size of 256.

In our research, the same dataset may be employed across multiple experiments. To optimize efficiency, we propose a one-time preprocessing step, during which the dataset is serialized to disk. This serialized representation allows seamless reuse in various experiments without redundant preprocessing. However, for the method-name-prediction experiment, an additional preprocessing step is necessary to extract method-names for each individual method. By adopting this approach, researchers can strike a balance between resource utilization and experimental flexibility, streamlining the overall research process.

\subsection{Code-Clone Detection}

Code-clone detection involves identifying code fragments that are similar or identical, in terms of syntax or semantics.
Code clones can be classified into four major types, as defined by Liu et al.~\cite{liuCanNeuralClone2021}:
\begin{itemize}
        \item
              Type-1: Code fragments that are identical except for variations in white space, layout, and comments. These are also known as exact or textual clones.

        \item
              Type-2: Code fragments that are identical except for variations in identifier names and literal values.
              These are also known as renamed or parameterized clones.

        \item
              Type-3: Code fragments that are syntactically similar but differ at the statement level.
              These are also known as gapped or near-miss clones.

        \item
              Type-4: Code fragments that are syntactically dissimilar but implement the same functionality.
              These are also known as semantic or functional clones.
\end{itemize}

In this paper, we propose a novel method for code-clone detection based on code normalization, code transformation, and contrastive learning.
We first normalize the code by removing comments and renaming variables, as described in Section \ref{sec_code_normalsize}.
This step can eliminate the syntactic variations that cause Type-1 and Type-2 clones.
Then, we extract short code paths from the normalized code using a tree-based traversal algorithm, as described in Section \ref{sec_code_extract}.
This step can capture the local structure and semantics of the code.
Finally, we use contrastive learning to learn code embeddings from the code paths, such that similar code paths have close embeddings and dissimilar code paths have distant embeddings.
This step can enable the detection of semantic clones, by measuring the similarity of the code embeddings.

We evaluated our method on two code-clone detection datasets:
POJ-104 \cite{zhangNovelNeuralSource2019} and BigCloneBench \cite{svajlenko2014towards, wang2020detecting}.
We compared our proposed method with the state-of-the-art pre-trained code representation model CodeBERT, and the unsupervised code-representation learning model InferCode.
POJ-104 contains code pairs, written in C, that are semantically equivalent but syntactically different \cite{zhangNovelNeuralSource2019}.
It contains 52,000 code fragments. 
BigCloneBench is a widely used benchmark dataset \cite{svajlenko2014towards, wang2020detecting}, comprising projects from 25,000 Java repositories:
It covers ten functionalities, and includes 6,000,000 true clone pairs and 260,000 false clone pairs.
Both datasets can be accessed from the CodeXGLUE GitHub repository\footnote{\textbf{CodeXGLUE}: \url{https://github.com/microsoft/CodeXGLUE}.}.

We built the OJClone dataset \cite{buiInferCodeSelfSupervisedLearning2021} following Bui et al. \cite{buiInferCodeSelfSupervisedLearning2021}, using code pairs from POJ-104 based on pairwise similarity.
500 programs from each of the first 15 POJ-104 problems were selected, resulting in 1.8 million clone pairs and 26.2 million non-clone pairs.
A comparison of all the pairs would be prohibitively time-consuming, so 50,000 clone pairs and 50,000 non-clone pairs were randomly selected for the code-detection evaluation.
Both the OJClone and BigCloneBench datasets were used to evaluate the performance, in both unsupervised and supervised settings.

\subsubsection{Unsupervised Learning
        \label{sec_exp_unsupervised}}

We first evaluated the model's performance in an unsupervised learning setting, where the model does not have access to the ground-truth labels during training, and relies on the contrastive loss to guide its learning.

We trained our framework's encoder with different parameter configurations on the Java-based BigCloneBench dataset.
This dataset is a benchmark dataset for clone detection, and contains nearly 901,000 clone pairs found within an inter-project Java source code dataset. The dataset is publicly available on GitHub and Hugging Face \footnote{\url{https://huggingface.co/datasets/code_x_glue_cc_clone_detection_big_clone_bench}.}.
We used unique training samples from the BigCloneBench training set. This is because the BigCloneBench training set data appears in paired format, indicating whether this is a cloned or non-cloned pair. This means that each sample may have multiple occurrences in a clone or a non-clone pair. Therefore, we filtered out duplicate samples from the BigCloneBench training set. Finally, we obtained 7302 unique samples for training. Figure \ref{fig_unique} in the Appendix shows how the pairs were formatted. We trained our proposed framework, which does not require any labels, on these samples. Our framework learns to distinguish between similar and dissimilar code snippets based on their embeddings. Our dataset settings for the other unsupervised tasks were also the same as for the BigCloneBench dataset.

In the validation phase, we computed the cosine similarity of the two ASTs and classified the code pairs as similar or dissimilar based on a threshold value.
In our experiments, we set the threshold value $T$ to 0.75.
This threshold value in code-clone detection is a hyperparameter that defines the cutoff point between similar and dissimilar code pairs based on their cosine similarity. We selected 0.75 as the threshold because it was the optimal value for our experimental settings to achieve high accuracy and recall. However, we recognize that the best threshold may change depending on the data features, the model structure, and the GPU resources. We also performed the experiment with a threshold of 0.5 ($T=0.5$) as shown in Table~\ref{tab:clone_ours}. 
The performance worsens and the training time increases when we reduce the threshold to 0.5. This is because our encoder has a simple architecture (2 or 4 layers) and our batch size is constrained by the GPU resources. Our model struggles to distinguish clones from non-clones with a lower threshold. Contrastive learning typically requires a large batch size, as it benefits from more negative samples during training. Therefore, we prefer a higher threshold ($T=0.75$) to enhance the discrimination.

\begin{table*}[!t]
        \centering
        \caption{Metrics for models with different architectures on BigCloneBench \(Java\)}
        \label{tab:clone_ours}
        \resizebox{\textwidth}{!}{
            \renewcommand{\arraystretch}{1.25}
                \begin{tabular}{rcrrrrrrrrrrr}
                        \toprule
                        \textbf{Model}     & \multicolumn{7}{c}{\textbf{Architecture \& Configuration}} & \multicolumn{4}{c}{\textbf{Metrics}}                                                                                                                                                                                                   \\ \midrule
                        \textbf{TransformCode} & \textbf{$n_{parameters}$}                                  & \textbf{$n_{gradients}$}             & \textbf{$n_{layers}$} & \textbf{$d_{model}$} & \textbf{$n_{heads}$} & \textbf{$d_{head}$} & \textbf{batch size} & \textbf{accuracy} & \textbf{precision} & \textbf{recall}  & \textbf{F1 Score} \\
                        \cmidrule{2-12}
                +relative pos -MLP ($T=0.75$) & 171,237,888                                                & 85,618,944                           & 4                     & 1024                 & 16                   & 64                  & 86                  & 85.67\%           & 80.56\%            & 85.67\%          & 81.51\%           \\
                \rowcolor{lightblue} +relative pos -MLP ($T=0.75$) & 137,621,760                                                & 68,810,880                           & 2                     & 1024                 & 16                   & 64                  & 128                 & 84.93\%           & 79.80\%            & 84.93\%          & 81.30\%           \\
                 +relative pos -MLP ($T=0.75$) & 122,149,376                                                & 61,074,688                           & 4                     & 768                  & 12                   & 64                  & 128                 & 85.74\%           & 80.60\%            & 85.74\%          & 81.48\%           \\
                \rowcolor{lightblue} +relative pos -MLP ($T=0.75$) & 100,076,800                                                & 50,038,400                           & 2                     & 768                  & 12                   & 64                  & 192                 & 84.58\%           & 79.81\%            & 84.58\%          & 81.35\%           \\
                +relative pos -MLP ($T=0.75$) & 77,255,168                                                 & 38,627,584                           & 4                     & 512                  & 8                    & 64                  & 192                 & 84.88\%           & 80.03\%            & 84.88\%          & 81.48\%           \\
                \rowcolor{lightblue} +relative pos -MLP ($T=0.75$) & 64,628,992                                                 & 32,314,496                           & 2                     & 512                  & 8                    & 64                  & 256                 & 79.53\%           & 78.66\%            & 79.53\%          & 79.08\%           \\
                +relative pos +MLP ($T=0.75$) & 79,095,808                                                 & 39,547,904                           & 4                     & 512                  & 8                    & 64                  & 192                 & 86.57\%           & 85.12\%            & 86.57\%          & 80.78\%           \\
                \rowcolor{lightblue} +relative pos +MLP ($T=0.75$) & 79,079,168                                                 & 39,539,584                           & 4                     & 512                  & 16                   & 32                  & 192                 & 86.44\%           & 82.43\%            & 86.44\%          & 81.47\%           \\
                +relative pos +MLP ($T=0.75$) & 126,089,728                                                & 63,044,864                           & 4                     & 768                  & 12                   & 64                  & 128                 & 86.59\%           & 84.52\%            & 86.59\%          & 80.95\%           \\
                \rowcolor{lightblue} +relative pos +MLP ($T=0.75$) & 144,448,256                                                & 72,224,128                           & 2                     & 1024                 & 16                   & 64                  & 128                 & \textbf{87.50\%}  & \textbf{84.76\%}   & \textbf{87.50\%} & \textbf{82.36\%}  \\
                \rowcolor{lightgrey} +relative pos +MLP ($T=0.5$) & 144,448,256                                                & 72,224,128                           & 2                     & 1024                 & 16                   & 64                  & 128                 & {80.34\%}  & {78.25\%}   & {80.34\%} & {79.24\%}  \\
                -relative pos -MLP ($T=0.75$) & 137,601,024                                                & 68,800,512                           & 2                     & 1024                 & 16                   & 64                  & 128                 & 41.19\%           & 76.09\%            & 41.19\%          & 48.57\%           \\
                \rowcolor{lightblue} -relative pos +MLP ($T=0.75$) & 144,427,520                                                & 72,213,760                           & 2                     & 1024                 & 16                   & 64                  & 128                 & 63.98\%           & 77.56\%            & 63.98\%          & 69.13\%           \\
                        \bottomrule
                \end{tabular}
        }
\end{table*}

We varied the encoding layers, hidden sizes ($d_{model}$), and the dimensions of the multi-head attention ($d_{head}$) of encoders of our framework.
We also examined how relative-positional encoding (+relative pos) and Multi-layer Projection (+MLP) influenced the model performance.
The experimental results in Table \ref{tab:clone_ours} show that relative-positional encoding is essential for the encoder to work well, and that MLP can moderately enhance the model performance.
We also found that increasing the model depth ($n_{layers}$) or hidden size ($d_{model}$) could improve the performance, as long as the dimensions of multi-head attention were fixed at 64 \cite{vaswani2017attention}.
Our framework achieved the best performance and convergence speed when using relative-positional encoding and MLP, two encoding layers, and a hidden size of 1024.
With these settings and a batch size of 128, it converged in less than 35 epochs and attained an F1 Score of 82.36\% and an Accuracy of 87.50\%.
To explore our framework's performance compared to other state-of-the-art models, we report the results of other unsupervised models on the same dataset in Table~\ref{tab:clone_compare} (adapted from Bui et al.~\cite{buiInferCodeSelfSupervisedLearning2021}).
All the models in Table~\ref{tab:clone_compare} were trained without labeled data.

\begin{table}[!b]
        \centering
        \caption{Metrics for different models on BigCloneBench \(Java\) (Unsupervised)}
        \label{tab:clone_compare}
        \resizebox{0.48\textwidth}{!}{
            \renewcommand{\arraystretch}{1.25}
        \begin{tabular}{rcrrr}
                \toprule
                \textbf{Model} & \multicolumn{3}{c}{\textbf{Metrics}}                                        \\ \midrule
                               & \textbf{Precision}                   & \textbf{Recall}  & \textbf{F1 Score} \\
                \cmidrule{2-4}
                Deckard        & 93.00\%                              & 2.00\%           & 3.00\%            \\
                DLC            & \textbf{95.00\%}                     & 1.00\%           & 1.00\%            \\
                SourcererCC    & 88.00\%                              & 2.00\%           & 3.00\%            \\
                Code2vec       & 82.00\%                              & 40.00\%          & 60.00\%           \\
                CodeBERT       & 77.48\%                              & 19.86\%          & 16.43\%           \\
                InferCode      & 90.00\%                              & 56.00\%          & 75.00\%           \\
        \rowcolor{lightblue} TransformCode ($n_{layers}=2,d_{model}=1024$)  & 84.76\%                              & \textbf{87.50\%} & \textbf{82.36\%}  \\
                \bottomrule
        \end{tabular}
        }
\end{table}

In this experiment, we focus on the unsupervised SE task of code-clone detection, and compare TransformCode with other unsupervised clone detectors that do not need labeled data. 
TransformCode can also perform supervised clone detection. 
We do not compare our framework with techniques that depend on supervised learning to build clone classifiers, such as: Oreo \cite{saini2018oreo}, CCD \cite{fangFunctionalCodeClone2020}, ASTNN \cite{zhangNovelNeuralSource2019}, and CCDLC \cite{CCDLC_2018,sheneamer_2017}. 
We also did not compare TransformCode with the work of Tufano et al.~\cite{10.1145/3196398.3196431}, who used a supervised learning technique to train a neural network that learns the semantic similarity between code components from a stream of identifiers. Our baselines for code-clone detection comparison were: 
Deckard \cite{jiang2007deckard}, SourcererCC \cite{sajnani2016sourcerercc}, DLC \cite{white2016deep}, Code2vec \cite{alon2019code2vec,kang2019assessing}, and InferCode \cite{buiInferCodeSelfSupervisedLearning2021}. To show the limitations of CodeBERT, which is only an encoder for code, we included it in the experiment using an unsupervised setting. 
Cosine similarity was used to measure the distance between two code embeddings from two code snippets, without any training. 
CodeBERT was not trained with a supervised clone-detection classifier, as this would have violated the unsupervised learning assumption.
Both Code2vec and InferCode use a similar prediction approach to ours, which is to predict the clone label based on the cosine similarity between two code snippets.
Table~\ref{tab:clone_compare} shows the results of the comparison.
Our framework achieves slightly lower Precision than InferCode, but it surpasses InferCode in terms of F1 Score and Recall. 
InferCode has lower Recall and higher Precision, which means it is less consistent in identifying code-clone pairs. 
However, our framework is more stable and reliable, as shown by its balanced Precision and Recall.
Our framework uses fewer model parameters and has faster training convergence compared to InferCode.

Table~\ref{tab:clone_ours} shows that the relative-position encoding and MLP improve our framework's performance.
Therefore, we applied this configuration to all encoders of our framework on the OJClone dataset. We omitted the AddTryCatch transformation in the AST transformation because C does not have it.
Due to our limited computing power (only two Nvidia RTX 3090 graphics cards), we conducted the experiments using only three different model parameter settings ($n_{layers}=2,d_{model}=1024$; 
$n_{layers}=4,d_{model}=1024$; and 
$n_{layers}=8,d_{model}=128$).

\begin{table}[!b]
        \centering
        \caption{Metrics for different models on OJClone \(C\) (Unsupervised)}
            \resizebox{0.48\textwidth}{!}{
            \renewcommand{\arraystretch}{1.25}
                \begin{tabular}{rcrrr}
                        \toprule
                        \textbf{Model}                       & \multicolumn{3}{c}{\textbf{Metrics}}                                        \\ \midrule
                                                             & \textbf{Precision}                   & \textbf{Recall}  & \textbf{F1 Score} \\
                        \cmidrule{2-4}
                        Deckard                              & 99.00\%                              & 5.00\%           & 10.00\%           \\
                        DLC                                  & 71.00\%                              & 0.00\%           & 0.00\%            \\
                        SourcererCC                          & 7.00\%                               & 74.00\%          & 14.00\%           \\
                        Code2vec                             & 56.00\%                              & 69.00\%          & 61.00\%           \\
                        CodeBERT                             & 77.48\%                              & 19.86\%          & 16.43\%           \\
                        InferCode                            & 61.00\%                              & \textbf{70.00\%} & 64.00\%           \\
                        \bottomrule
                        TransformCode ($n_{layers}=2,d_{model}=1024$) & 67.14\%                              & 64.68\%          & 63.36\%           \\
            \rowcolor{lightblue}    TransformCode ($n_{layers}=4,d_{model}=1024$) & \textbf{67.69\%}                     & 67.29\%          & \textbf{67.10\%}  \\
                        TransformCode ($n_{layers}=8,d_{model}=128$)  & 65.52\%                              & 64.04\%          & 63.16\%           \\
                        \bottomrule
                \end{tabular}
        }
        \label{tab:ojclone_compare}
\end{table}

Our framework outperforms the benchmark models and achieves comparable results to InferCode, the state-of-the-art model, as shown in Table \ref{tab:ojclone_compare}.
Our proposed model attains a higher precision than InferCode under three different parameter settings:
two layers with a hidden size of 1024; 
four layers with a hidden size of 1024; and 
eight layers with a hidden size of 128.
With four layers and a hidden size of 1024, our proposed model achieves the best performance with 67.69\% precision and 67.10\% F1 Score.
Using an 8-layer encoder with a hidden size of 128, it has poorer performance than those with hidden sizes of 1024 because of the smaller hidden size for encoding.
Moreover, our framework outperforms the other models under all three different parameter settings.
These results demonstrate the effectiveness of our framework in learning code representations and detecting code clones.

We conducted experiments on two datasets in different programming languages and again demonstrated the effectiveness of our approach.
Our approach relies on unsupervised comparison learning, which requires a high-quality anchor AST.
To enhance the model's generalization and robustness, we will consider various additional types of AST transformations in the future.

\subsubsection{Visualization for Unsupervised Learning
        \label{sec_visualization}}

\begin{figure*}[!t]
        \centering
        \graphicspath{{img/}}
        \subfigure[Visualization of the Code Embedding OJClone C dataset]{
                \includegraphics[width=0.48\textwidth]{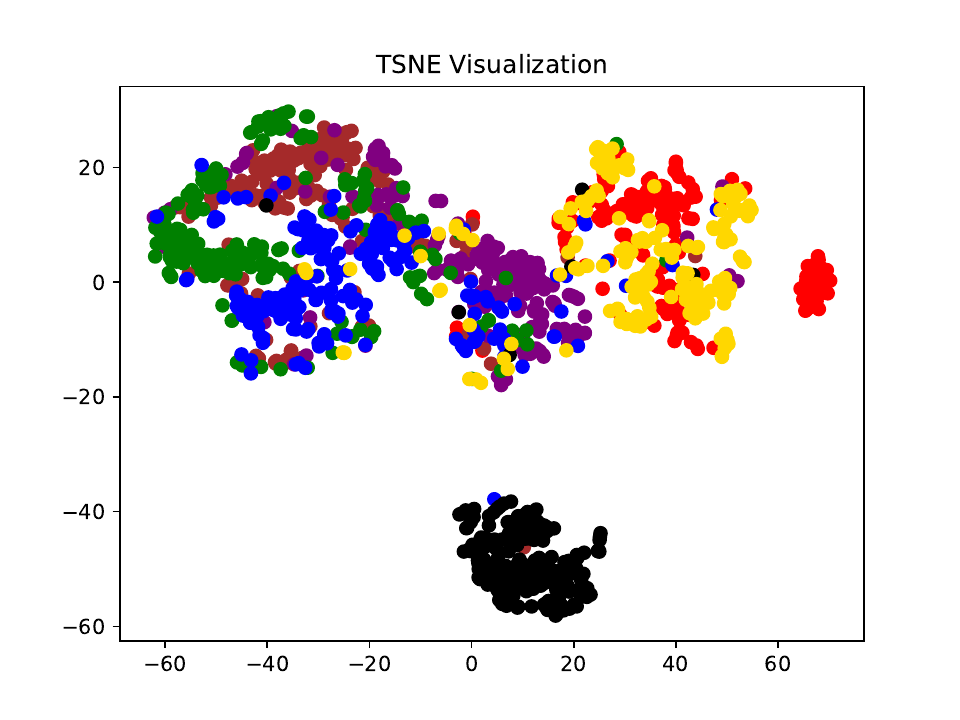}
        }
        \subfigure[Visualization of the Code Embedding in Java Function Dataset]{
                \includegraphics[width=0.48\textwidth]{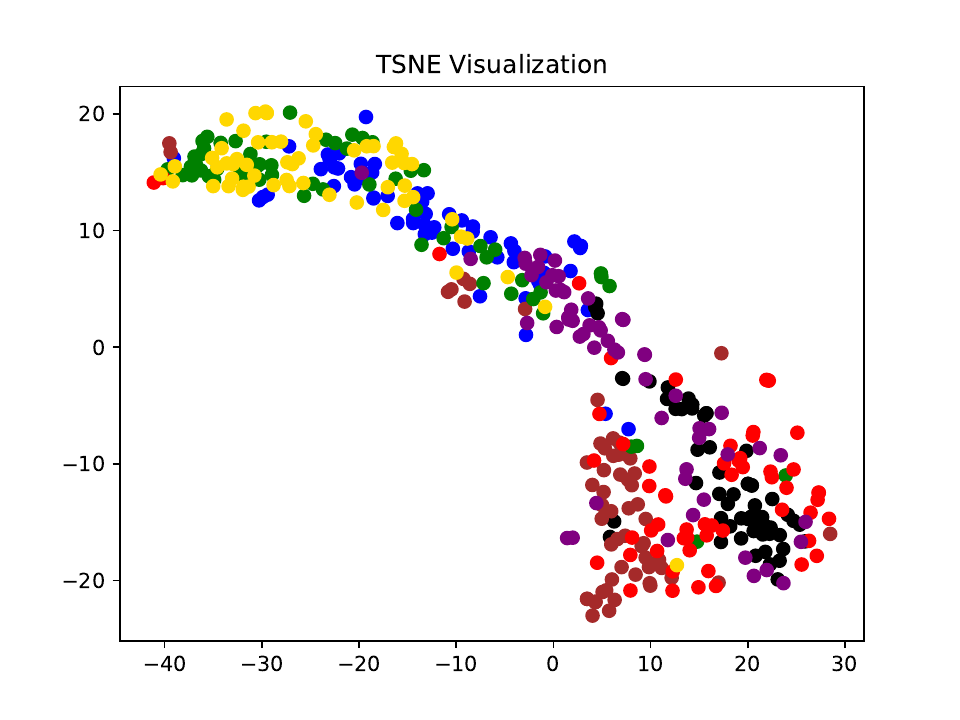}
        }
        \caption{Visualization with TSNE for two different datasets.}
        \label{fig_tsne}
\end{figure*}

To help better understand what our proposed model learns from the code, we created a visualization of
the code embedding from our proposed model using two layers with hidden sizes of 1024.
The code embedding is a high-dimensional representation of the syntactic and semantic features of the code snippets, which can capture the similarities and differences among them.
TSNE is a popular technique for dimensionality reduction and visualization of high-dimensional data.
It preserves the local structure and distances of the data points.
We used TSNE \cite{JMLR:v9:vandermaaten08a} to reduce 1024-dimensional code vectors into two-dimensional vector space, and then plotted them using matplotlib \cite{matplotlib_Hunter_2007}:
The results are shown in Fig.~\ref{fig_tsne}.
The plots show that our framework can learn to group similar code into the same cluster, even though we use contrastive learning with a relatively small batch size of 86.
This indicates that our framework can effectively learn from the code snippets without requiring a large amount of data.
One of the challenges for our approach is the GPU computation and memory limitation, which affects the quality of the group clusters and the code-embedding boundaries.
Training with a larger batch size could improve this aspect.
Nevertheless, our method has the advantage of
converging with small epochs of training and being adaptable to any custom dataset.
This gives our framework flexibility and scalability for various code-analysis tasks.

\subsubsection{Supervised Learning
        \label{sec_supervised_learn}}

In the supervised-learning experiment, we evaluated the performance of our proposed model (and the other models) for clone detection in a supervised learning setting.
Labeled data from the OJClone C dataset was used to train the models.
In this experiment, our model consists of a transformer encoder with relative positional encoding and a linear classifier, which were jointly trained with cross-entropy loss as the supervised loss.
The encoder also employs contrastive loss, as explained in Section \ref{sec_framework}.
We used a trainable parameter $\alpha = 0.2$ as the initial value to balance the two losses with the following formula:
\begin{equation}
        \small
        Loss = \alpha * Loss_{Contrastive} + (1 - \alpha) * Loss_{Supervised}.
\end{equation}
The transformed AST was used as a data augmentation technique to enhance the supervised learning performance.

We compared our framework with several baselines, including traditional methods such as SVMs, neural network models (TextCNN \cite{kimConvolutionalNeuralNetworks2014}, LSTM \cite{zarembaLearningExecute2015}, TBCNN \cite{mouConvolutionalNeuralNetworks2015}, and LSCNN \cite{huoEnhancingUnifiedFeatures2017a}), program dependency graphs-based (PDG-based) methods \cite{allamanisLearningRepresentPrograms2022,tufanoDeepLearningSimilarities2018} and CodeBERT \cite{codebert2020}.
In addition, we conducted an ablation study utilizing the same transformer encoder as in TransformCode, coupled with a linear classifier for comparison. This encoder model underwent training without any data augmentation through AST transformation.

We used linear kernels for the SVMs, and extracted statistical features from TF-IDF, N-gram, and LDA.
We set the N-gram size to 2, and limited the maximum number of features to 20,000.

The code was treated as plain text for TextCNN and LSTM, both of which used the same hyperparameters.
The TextCNN kernel size was 3, and the number of filters was 100.
The hidden state dimension for LSTM was 100.

The PDG-based graph embedding involved the construction of program graphs from PDGs, and the application of two graph-embedding methods, HOPE \cite{ouAsymmetricTransitivityPreserving2016} and Gated Graph Neural Networks \cite{liGatedGraphSequence2017}.
Incomplete code snippets had header files added to enable their PDG construction.
The final max pooling layer was used to get the code embeddings.

To adapt the encoder-based pre-trained model CodeBERT to this experiment, we appended feed-forward and classifier layers after CodeBERT. 
These additional layers were trained with the same hardware configuration.

\begin{table}[!b]
        \centering
        \caption{Metrics for model trained on OJClone \(C\) (Supervised)}
        \label{tab:supervised_result}
        \resizebox{0.48\textwidth}{!}{
            \renewcommand{\arraystretch}{1.25}
        \begin{tabular}{rcrcrcr}
                \toprule
                \textbf{Group}                              & \phantom{abc} & \multirow{1}{1.5cm}{\textbf{Methods}} & \textbf{Accuracy} \\
                \midrule
                \multirow{3}{0.9cm}{\textbf{SVMs}}          &               & SVM+IF-IDF                            & 79.4\%            \\
                                                            &               & SVM+N-gram                            & 84.7\%            \\
                                                            &               & SVM+LDA                               & 47.9\%            \\ \midrule
                \multirow{7}{0.9cm}{\textbf{Neural Models}} &               & TextCNN                               & 88.7\%            \\
                                                            &               & LSTM                                  & 88.0\%            \\
                                                            &               & TBCNN                                 & \textbf{94.0\%}   \\
                                                            &               & LSCNN                                 & 90.9\%            \\
                                                            &               & PDG+HOPE                              & 4.2\%             \\
                                                            &               & PDG+GGNN                              & 79.6\%            \\
                                                            &               & CodeBERT                              & 93.3\%            \\
\rowcolor{lightblue}                                        &               & TransformCode ($n_{layers}=2,d_{model}=128$)   & 93.5\%            \\
\rowcolor{lightgrey}  &                                                     & Transformer Encoder ($n_{layers}=2,d_{model}=128$) & 93.1\%            \\
                \bottomrule
        \end{tabular}
        }
\end{table}

Table~\ref{tab:supervised_result} shows that our framework achieves comparable performance to TBCNN (the best performer), but has fewer parameters, demonstrating the effectiveness of our method.
Employing AST-based data augmentation has marginally enhanced the efficacy of our framework in the supervised learning experiments. Importantly, it serves as an integral component for generating anchor samples for contrastive learning in our framework.
Our framework also achieved slightly better results than CodeBERT, with a much smaller model size, and fewer parameters:
We only used two encoder layers and 128 hidden units for the encoder, while CodeBERT uses 12 encoder layers and 768 hidden units. 
This demonstrates the efficiency of our framework in learning from code data.

\subsection{Method-Name Prediction}

Code2vec \cite{alon2019code2vec} and Code2seq \cite{alon2018code2seq} are  code-embedding models that use method-name prediction as the objective function to learn semantic representations of code snippets.
They were both trained using a Java-Small dataset, which is a collection of Java methods extracted from open-source projects.
However, Code2vec and Code2seq have some limitations, such as being sensitive to syntactic variations, and being unable to handle OOV tokens, because they lack code normalization.
We compared the performance of our framework with InferCode \cite{buiInferCodeSelfSupervisedLearning2021}, Code2vec, and Code2seq on the Java-Small dataset, finding that our framework outperforms them according to several metrics.

More details about the Java-Small dataset and the method-name prediction task are as follows:
\begin{itemize}
        \item
              The Java-Small dataset consists of three parts:
              training;
              testing; and
              validation.
              Each part contains Java methods from different projects, and each method is associated with a method name as the label. The dataset is preprocessed by splitting the method names into subwords based on camel case and punctuation, and by extracting the AST paths from the code snippets.
              After preprocessing, there were 73,998 methods in the training set;
              4,818 methods in the testing set; and
              1,702 methods in the validation set.

        \item
              The method-name prediction task is an SE task that aims to generate a descriptive and concise name for a given code snippet.
              It is a challenging task that requires understanding the functionality and semantics of the code snippet.
              The task can be formulated as a sequence-generation problem, where the input is a code snippet and the output is a sequence of subwords that form the method name.
              Since the method names are usually short, we limit the output sequence length to five subwords, 
              which is consistent with  Code2vec's output sequence length. 

        \item
              To evaluate the performance of our method-name prediction model, we used the same evaluation metrics as in Code2vec \cite{alon2019code2vec} and InferCode \cite{buiInferCodeSelfSupervisedLearning2021}, which used the F1 Score of the sub\-words of the generated method name.
              A sub\-word is a meaningful part of a word that is separated by camel case, underscore, or hyphen.
              For example, the word computeMax has two sub\-words: compute and Max.
              We compared the sub\-words of the predicted method name with the sub\-words of the ground-truth method name and calculated the Precision, Recall, and F1 Score, accordingly.
              For instance:
              If a predicted method name is \textit{compute\_max}, it will be considered as an exact match to the ground-truth method name \textit{computeMax}, and the F1 Score will be 100\%.
              If the predicted result is \textit{getMax}, then the sub\-word \textit{Max} will get full precision but only a 50\% score for Recall, and the F1 Score would be 67\%.
              If the predicted result is \textit{getMaxResult}, it will get full Recall but only 67\% Precision, and the F1 Score will be around 80\%.
\end{itemize}

We encoded the source code using the structure described Section \ref{sec_framework}, which consists of a code tokenizer and an encoder. 
We also used an additional method name tokenizer that only tokenizes the method name.

To enhance the data quality and diversity, we applied the AST transformation techniques from Section~\ref{sec_code_to_ast}.
These techniques preserve the functionality and semantics of the original code snippet, but change its surface form:
This makes it possible to generate more variations of code snippets for one method, while still using the same method name as the label.
Finally, the code tokenizer converts the extracted AST path into a sequence of tokens, and the encoder transforms the tokens into vector representations.
We attached a one-layer LSTM decoder to our encoder to output the method name. 
The LSTM decoder was an RNN that can generate output sequences from a fixed-length vector representation of the input.
We removed special tokens (such as $<bos>$ and $<eos>$) from the decoder output, and only predicted five words at most as our final result.

We also employed the same transformer encoder as presented in TransformCode in this experiment, appended with the same one-layer LSTM decoder to facilitate ablation studies. Notably, this model was not trained with any data augmentation through AST transformation.
Because CodeBERT does not have a decoder component, we combined it with an LSTM decoder that was identical to the one used in our proposed method.

For comparison purposes, we used the same one-layer LSTM-decoder across three different models in this experiment: CodeBERT, TransformCode, and a Transformer Encoder. 
Our experimental setup was carefully controlled to minimize the influence of extraneous variables on performance outcomes, ensuring a fair comparison under identical hardware configurations. We acknowledge that integrating more sophisticated decoder layers could potentially enhance performance; however, such modifications are constrained by the computational limits of our available hardware (specifically the RTX 3090 GPU). Notably, CodeBERT requires more computational resources than TransformCode, and the addition of a complex decoder could lead to out-of-memory (OOM) errors with CodeBERT. As a result, our primary objective was to showcase the effectiveness of our framework, considering the same hardware resources.

\begin{table}[!t]
        \centering
        \caption{Result of method name prediction in F1 Score on Java-Small dataset (Supervised)}
        \label{tab:method_java_small}
        \setlength{\tabcolsep}{5mm}
        \resizebox{0.48\textwidth}{!}{
            \renewcommand{\arraystretch}{1.25}
        \begin{tabular}{rcrr}
                \toprule
                \textbf{Model} & \multicolumn{2}{c}{\textbf{Supervised}}                     \\
                \midrule
                               &                                         & \textbf{F1 Score} \\
                \cmidrule{2-3}
                InferCode      &                                         & 35.67\%           \\
                Code2vec       &                                         & 18.62\%           \\
                Code2seq       &                                         & \textbf{43.02\%}           \\
                CodeBERT       &                                         & 19.98\%           \\
\rowcolor{lightblue} TransformCode ($n_{layers}=2,d_{model}=512$)  &                                         & 34.80\%  \\
\rowcolor{lightgrey} Transformer Encoder ($n_{layers}=2,d_{model}=512$)  &                                         & 32.10\%  \\
                \bottomrule
        \end{tabular}
        }
\end{table}

Table~\ref{tab:method_java_small} shows the results of our framework on the Java-Small dataset for method-name prediction.
As shown in Table~\ref{tab:method_java_small}, our framework achieves comparable performance to InferCode, in terms of F1 Score. 
It is important to note the difference in the number of parameters in our framework's encoder and CodeBERT, as well as the differences in their respective training datasets (CodeBERT was pre-trained on a different training set). Consequently, we did not fine-tune CodeBERT but rather opted to train only the LSTM decoder, maintaining the same hardware setup. 
Notably, CodeBERT has a larger dictionary, while our framework's dictionary was tailored specifically for this task using only training data, without access to test data. 
The dictionary size for our framework in this experiment was approximately 10,000 entries.
Despite these differences, our framework was more efficient than CodeBERT, as it used a smaller encoder and had a faster training speed. 
Our framework could be trained around four times faster than CodeBERT on the same hardware configuration.
Furthermore, leveraging AST data augmentation slightly enhanced the performance of our framework in this supervised learning experiment.

\subsection{Code Classification}

The POJ-104 dataset is a large-scale collection of programming problems and solutions.
It contains a total of 52,000 programs in 104 categories, covering different topics and difficulty levels.
To evaluate the performance of our framework, we divided the dataset into three subsets (in a ratio of 7:2:1):
training;
testing; and
validation.
The training set was used to train the model parameters, the testing set was used to measure the generalizability of the model, and the validation set was used to tune the hyperparameters and prevent overfitting.

Our framework consists of several components, including an AST encoder, a category classifier, and a code Transformer.
The category classifier predicts the category of the code snippet from its code vector using a fully connected layer and a softmax layer.
We also employed the same transformer encoder as in TransformCode in this experiment as an ablation study. This model was not trained with any data augmentation through AST transformation.

The goal of our framework is to learn a code-vector representation that is invariant to the code transformations and discriminative for the code categories.
To achieve this goal, we used a combination of contrastive loss and cross-entropy loss as our training loss function.
Contrastive loss measures the similarity between the original and transformed code vectors, and encourages them to be closer in the latent space.
Cross-entropy loss measures the accuracy of the category prediction for both the original and transformed codes, and encourages them to have high confidence scores for their true categories.
We used a trainable parameter $\alpha$ to balance the two losses, initializing it with a value of 0.1.
The loss function was defined as follows:
\begin{equation}
        \small
        \begin{aligned}
                Loss & = \alpha * Loss_{Contrastive} + (1 - \alpha) * Loss_{Category} \\
                     & + 0.5* Loss_{Anchor},
        \end{aligned}
\end{equation}
where $Loss_{Contrastive}$ is the contrastive loss between the original and transformed code vectors, calculated using a cosine similarity function;
$Loss_{Category}$ is the cross entropy loss for the category prediction of the original code, calculated with a softmax function over the output of the category classifier;
$Loss_{Anchor}$ is the cross entropy loss for the category prediction of the transformed code, calculated using the same softmax function; and
$\alpha$ is a trainable parameter that controls the trade-off between the contrastive loss and the category loss.
The coefficient of 0.5 for $Loss_{Anchor}$ is a hyperparameter that can be adjusted based on different datasets
---
we determined the value through empirical examination.
Both the original code and the transformed code should have the same category label, which we used as the ground truth for both $Loss_{Category}$ and $Loss_{Anchor}$.

\begin{table}[!t]
        \centering
        \caption{Result of code classification in Accuracy on OJ C dataset (Supervised)}
        \label{tab:method_oj}
        \setlength{\tabcolsep}{5mm}
        \resizebox{0.48\textwidth}{!}{
            \renewcommand{\arraystretch}{1.25}
        \begin{tabular}{rcrr}
                \toprule
                \textbf{Model} & \multicolumn{2}{c}{\textbf{Supervised}}                     \\
                \midrule
                               &                                         & \textbf{Accuracy} \\
                \cmidrule{2-3}
                InferCode      &                                         & 94.00\%           \\
                TextCNN        &                                         & 88.70\%           \\
                Bi-LSTM        &                                         & 88.00\%           \\
                ASTNN          &                                         & \textbf{97.80\%}  \\
                CodeBERT       &                                         & 73.70\%           \\
\rowcolor{lightblue}  TransformCode ($n_{layers}=2,d_{model}=1024$)  &                                         & 92.10\%           \\
\rowcolor{lightgrey}  Transformer Encoder ($n_{layers}=2,d_{model}=1024$)  &                                         & 91.08\%           \\
                \bottomrule
        \end{tabular}
        }
\end{table}

Table~\ref{tab:method_oj} shows that our framework achieved comparable results to two state-of-the-art models: InferCode and ASTNN.
In addition to the comparable performance, our framework has the advantage of being easy to train and requiring less GPU power.
Compared with ASTNN, which uses a tree-based network to encode the AST, our framework uses a token-based approach, which can capture more fine-grained information from the AST.
To adapt CodeBERT to this experiment, we appended feed-forward and classifier layers after CodeBERT. 
These additional layers were trained independently, as fine-tuning CodeBERT under the same hardware configuration was not feasible. Remarkably, our framework surpassed CodeBERT in terms of effectiveness using the identical hardware setup. Furthermore, owing to the incorporation of data augmentation, our framework exhibited a slight improvement over the transformer encoder --- the same encoder in our TransformCode.
Even though both are supervised tasks, this experiment is more challenging than clone detection, as it involves 104 categories.
Our future work will include combining the tree network and the token-based approach to explore their synergies for code-representation learning.

\subsection{Limitations and Threats to Validity}

One of the challenges we faced in our experiments was the limitation of GPU memory, which prevented us from using a large batch size for training.
This is a significant drawback, as contrastive learning methods often achieve higher performance when using large batch sizes.
However, we only had access to two Nvidia RTX 3090 graphics cards, which had a total of 48GB of GPU memory.
Therefore, we could not train a large model like CodeBERT, which has 12 layers and 768 hidden sizes, nor could we use a large batch size.
As a result, our proposed model might not have reached its full potential, and could benefit from further optimization.
In theory, training our proposed model with a larger batch size and more 
layers could lead to much better performance and more robust representations.

Another challenge we faced in our research was the diversity of programming languages and their syntax rules. For example, Java has more syntactic components than C, such as the try-catch statement, which allows for more options for code transformation.
Therefore, our proposed model may achieve better results with programming languages that have more structures, as they can generate more semantically equivalent variants for training.
Our future work will include consideration of different optimization levels when compiling the code as part of the transformation method:
This may be able to provide more reliable and consistent variants.
Using obfuscators to perform the transformation \cite{Obfuscators4Compilers} will also be explored.
We hypothesize that using more semantically equivalent variants for training will result in better performance of our proposed model, including learning more robust representations.
We look forward to exploring this further in our future work.

\section{Conclusions and Future Work}
\label{sec_conclusion}

In this paper, we have introduced a novel framework that learns code representations in an unsupervised manner, by applying contrastive learning and data-augmentation techniques to ASTs.
Our framework leverages attention mechanisms to capture the structural and semantic features of code-fragment ASTs across different programming languages, without requiring any labels.
To generate positive and negative samples for contrastive learning, we created anchor samples and code variants from the ASTs by applying various AST transformations.
We evaluated our framework on several datasets, and for three SE tasks:
code-clone detection, code classification, and method-name prediction.
The results from our evaluation confirm the superiority of our framework over many existing approaches.
For the task of unsupervised code-clone detection, our framework surpassed 13 existing approaches, including some well-known ones such as InferCode and Code2vec. 
Our framework is also encoder-agnostic and language-agnostic, which enables it to be easily applied to any other dataset, and to adjust different neural network parameters according to different computing resources. 
Moreover, our framework can be easily adapted for supervised learning tasks in SE by incorporating labels into the contrastive loss.
We have performed ablation studies to compare the efficacy of our proposed framework against the baseline encoder within a supervised learning context. The findings indicate that, by incorporating AST augmentation, our framework can achieve a marginal enhancement in performance.

In our future work, we plan to integrate large language models (LLMs) into our framework to address more practical and challenging SE problems, such as large-scale code retrieval and code-defect detection.
These tasks require a deep understanding of the semantics and syntax of code and the ability to generate or match code with natural language. 
We hope to leverage the power and knowledge of LLMs, especially those trained in large amounts of code, to enhance our performance of code embedding. 
We believe that LLMs can provide rich and diverse code representations that capture the code's syntactic and semantic information. 
We look forward to these investigations, and to being able to share our further findings with the community.

\section*{Acknowledgments}
We would like to thank the anonymous reviewers for their many constructive comments. This work is supported by the Science and Technology Development Fund of Macau, Macau SAR, under Grant Nos. 0021/2023/RIA1 and 0046/2021/A, and a Faculty Research Grant of Macau University of Science and Technology under Grant No. FRG-22-103-FIE. This work is also partially supported by the National Natural Science Foundation of China, under Grant Nos. 61872167 and 61502205.

~\\

\bibliographystyle{IEEEtran}
\bibliography{IEEEabrv,references}

\clearpage
\newpage
\setcounter{page}{1}
\pagenumbering{arabic}

\begin{figure*}[!t]
\parbox{\textwidth}{

\begin{center}
    \Huge{Appendix: TransformCode: A Contrastive Learning Framework for Code Embedding via Subtree Transformation}
    \vspace{10pt}
    
    \normalsize{Zixiang Xian, Rubing Huang, Dave Towey, Chunrong Fang, Zhenyu Chen}
    \vspace{10pt}
\end{center}
}

\parbox{\textwidth}{
\appendices
Figure \ref{fig_unique} shows how the pairs were formatted in the BigCloneBench dataset for unsupervised code-clone detection. We used 7302 unique training samples from the BigCloneBench dataset, which were formatted in pairs to include both clones and non-clones. Our TransformCode framework was trained on the unique samples, and the same dataset settings were applied to other unsupervised tasks, maintaining consistency with the BigCloneBench dataset.
}
\vspace{10pt}

\end{figure*}

 \setcounter{figure}{0}
 \renewcommand{\thefigure}{Appendix.\arabic{figure}}
 \renewcommand{\thefigure}{A.\arabic{figure}}
 \addtocounter{figure}{0}

\begin{figure*}[!t]
       \caption{The clone pair format for BigCloneBench.}
        \label{fig_unique}
        \centering
        \graphicspath{{img/}}
        \includegraphics[width=0.95\textwidth]{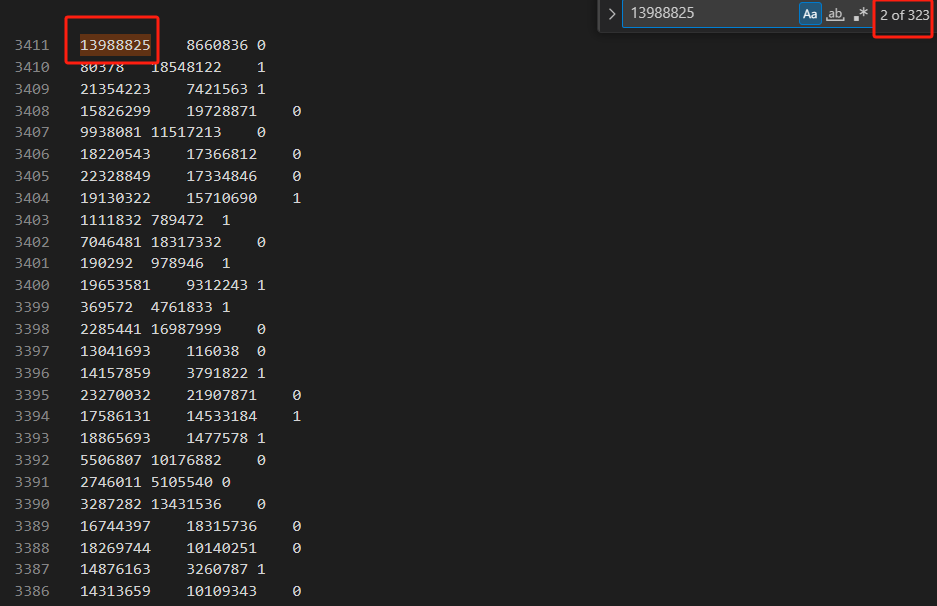}
\end{figure*}

\newpage

\vfill

\end{document}